\documentclass[12pt]{article}
\pdfoutput=1
\usepackage[utf8]{inputenc}
\usepackage{geometry}
\usepackage{titlesec}
\usepackage{graphicx}
\usepackage{caption}
\usepackage{bbold}
\usepackage{subcaption}
\usepackage{bigints}
\usepackage{booktabs}
\usepackage{physics}
\usepackage{slashed}
\usepackage[]{amsmath,amssymb,lmodern}
\usepackage{hyperref}
\usepackage[T1]{fontenc} 
\usepackage{soul}
\usepackage{dsfont}
\usepackage[dvipsnames]{xcolor}
\usepackage{mwe}
\usepackage{tikz}
\usepackage{mathtools}
\usepackage{empheq}
\usepackage{cite}

\newcommand\half{{1\over 2}}

\DeclareMathOperator{\im}{Im}

\def\nspc{\!\spc}
\def\eea{\end{eqnarray}}
\def\bea{\begin{eqnarray}}
\def\spc{\hspace{1pt}}
\def\is{\!\!&\! = \! & \!\!}

\def\tildeT{\widehat{T}}
\def\nn{\nonumber}
\def\p{\partial}

\def\ttau{\sigma}
\def\ssigma{\tau}
\def\({\left(}
\def\){\right)}
\def\hatK{{\cal K}}
\def\hatF{{F}}
\def\hatZ{{Z}}

\definecolor{vert}{rgb}{0.1367 0.543 0.1367}

\makeatletter\@addtoreset{equation}{section}\makeatother

\renewcommand{\title}[1]{\vbox{\center\LARGE{#1}}\vspace{5mm}}
\renewcommand{\author}[1]{\vbox{\center#1}\vspace{5mm}}

\newcommand{\email}[1]{\vbox{\center\tt#1}\vspace{5mm}}
	
\def\be{\bea}
\def\ee{\eea}

\def\Z{\mathbb{Z}}
\def\cH{{\cal H}}

\def\i{{\rm i}}

\begin{document}

	\begin{titlepage}
		\setcounter{page}{1} \baselineskip=15.5pt \thispagestyle{empty}

	  \begin{flushright}
	 \hfill{\tt CALT-TH 2022-039, PUPT-2637}
  \end{flushright} 

\begin{center}

 \hfill \\
 \hfill \\

\title{S-duality in $T\bar{T}$-deformed CFT
% $T\bar{T}, T^2$ and $T$-duality
}

~\vskip 0.01 in

\author{
	\resizebox{\textwidth}{!}{Nathan Benjamin$^a$, Scott Collier$^{b}$, Jorrit Kruthoff$^{c}$, Herman Verlinde$^{d}$, Mengyang Zhang$^{d}$}
}
~\vskip 0.05 in

\emph{\small $^{a}$Walter Burke Institute for Theoretical Physics, Caltech, Pasadena, CA 91125, USA}
%~\vskip .2 in
\emph{\small $^{b}$Princeton Center for Theoretical Science, Princeton University, Princeton, NJ 08544, USA}
%~\vskip .2 in
\emph{\small $^{c}$School of Natural Sciences, Institute for Advanced Study, Princeton, NJ 08540, USA}
%~\vskip .2 in 
\emph{\small $^{d}$Joseph Henry Laboratories, Princeton University, Princeton, NJ 08544, USA}
~\vskip .5 in 
\email{
	nbenjami@caltech.edu, scott.collier@princeton.edu, \\ kruthoff@ias.edu, verlinde@princeton.edu, mengyang@princeton.edu 
}

\end{center}

~\vskip .2 in 

\abstract{$T\bar{T}$ deformed conformal field theories can be reformulated as worldsheet theories of non-critical strings. We use this correspondence to compute and study the $T\bar{T}$ deformed partition sum of a symmetric product CFT. We find that it takes the form of a partition sum of a second quantized string theory with a worldsheet given by the product of the seed CFT and a gaussian sigma model with the two-torus as target space. We show that deformed symmetric product theory admits a natural UV completion that exhibits a strong weak coupling $\mathbb{Z}_2$ duality that interchanges the momentum and winding numbers and maps the $T\bar{T}$-coupling $\lambda$ to its inverse $1/\lambda$. The $\mathbb{Z}_2$ duality is part of a full O$(2,2,\mathbb{Z})$-duality group that includes a PSL$(2,\mathbb{Z})$ acting on the complexified $T\bar{T}$ coupling. The duality symmetry eliminates the appearance of complex energies at strong coupling for all seed CFTs with central charge $c\leq 6$. }

\vfill 

\noindent February 19, 2023 

\end{titlepage}

\renewcommand\Large{\fontsize{14.5}{14}\selectfont}
\renewcommand\large{\fontsize{13}{13}\selectfont}
\newpage
\tableofcontents
\thispagestyle{empty}
\newpage
\setcounter{page}{1}
	
\addtolength{\abovedisplayskip}{1mm}
\addtolength{\belowdisplayskip}{1mm}
\addtolength{\parskip}{1mm}
\section{Introduction}
	
Consider a two-dimensional conformal field theory with a discrete spectrum of states with energies $E_i = h_i+\bar{h}_i-\frac{c}{12}$ and momenta $j_i = h_i-\bar{h}_i \in \mathbb{Z}$. The $T \bar T$ deformation of a 2d conformal field theory is defined via the flow equation
\begin{eqnarray}
		S_{\rm QFT}(\lambda) \is  S_{\rm CFT} -\int\! d^2z \, O_{T\bar{T}}(\lambda) ,
		\qquad \ \nn \\[-2mm]\\[-2mm]\nn
		\partial_\lambda O_{T\bar{T}}(\lambda) \!\!\!\!\!\!\!\!\!\!\!\!\!\!\!\!& &\ \  \ =
		\, T\spc \bar{T} - \Theta^2
\end{eqnarray} 
with $\Theta$ the trace of the stress tensor and where $\lambda$ labels the $T\bar{T}$ coupling. This deformation has received considerable attention in recent years as a prime example of an exactly soluble irrelevant deformation that preserves the integrability properties of the undeformed CFT \cite{SMIRNOV2017363,Zamolodchikov:2004ce}.  As shown by Smirnov and Zamolodchikov \cite{SMIRNOV2017363}, the energies $\mathcal{E}_i(\lambda)$ of the individual energy eigenstates of the $T\bar{T}$ deformed theory depend in a universal way on the energies $E_i$ and momenta $j_i\in \Z$ of the corresponding states in the undeformed theory via
\bea
\label{edeform} \mathcal{E}_i(\lambda) \is  \frac{1}{\lambda}\bigl(-1+\sqrt{1+2E_i\lambda+ j_i^2\lambda^2}\bigr).
\eea
The energies ${\cal E}_i(\lambda)$ are all real provided the coupling is restricted to the range $\lambda \leq \frac 6c$.

In this paper we will compute and study the $T\Bar{T}$ deformed  partition function of $N$-fold symmetric product CFTs. $T\bar{T}$-deformations of symmetric product CFTs come in two types: single- or double-trace (and with various applications) \cite{Hashimoto:2019ttbar, Hashimoto:2019ttbarsym, McGough:2016lol, Callebaut:2019omt, Cavaglia:2016oda, Giveon:2017nie, Giveon:2017myj, Cardy:2018sdv, Dubovsky:2018bmo, LeFloch:2019rut, Dubovsky:2012wk, Dubovsky:2017cnj, Taylor:2018xcy, Hartman:2018tkw, Gross:2019ach, Gorbenko:2018oov, Cardy:2019qao, Cardy:2022mhn,Apolo:2019zai,Apolo:2023aho}. We will focus on the single-trace deformation. 	

Our first object of study is the grand canonical partition function of a $T\bar{T}$ deformed symmetric product CFT given by the weighted sum 
\bea
\label{symsum}
	\hatZ_{+}(\rho,\ssigma)  \is 1 + \sum_{N=1}^\infty \, p^N 
	Z_N(\rho_2, \ssigma), \qquad \quad p \equiv e^{2\pi \i \rho}, \ \lambda \equiv  \frac{\ssigma_2}{\rho_2}
\eea
with $Z_N(\rho_2, \ssigma)$ the deformed partition function of the $N$-fold symmetric product CFT 
\bea
	S^N {\rm CFT} = {\rm CFT}^N/S_N .
\eea 
Here $\ssigma$ denotes the modular parameter of the torus and  $\rho = \rho_1 + \i \rho_2$ is a complex parameter that encodes the $T\bar{T}$ coupling $\lambda$ and defines a fugacity parameter~$p$ that keeps track of the order $N$ of the symmetric product.  As we will make more explicit in what follows, $Z_{+}(\rho,\ssigma)$ is related to the partition function of a second quantized string theory with a worldsheet theory equal to the product of the seed CFT  with partition sum $Z_1(0,\ssigma)$ times a sigma model with a two-torus with modular parameter $\ssigma$ as target space and complexified volume modulus $\rho$, where $\rho_1$ defines the $B$-field flux through the target torus. This reformulation of the $T\bar{T}$ deformed symmetric product CFT will play a key role in what follows.
	 	 
Our formula for $Z_{+}(\rho,\ssigma)$, given in equation  \eqref{zttt}, takes the form of a non-chiral generalization of the DMVV formula \cite{Dijkgraaf:1996elliptic} with the energies $E_i$ replaced by the corresponding deformed energies  $\mathcal{E}_i(\lambda)$\footnote{The expression for $Z_{+}$ was derived in a similar way from modular invariance considerations in \cite{Apolo:2023aho}, where the large-$N$ behavior of the single-trace $T\bar{T}$-deformed partition function was studied.}. 
Motivated by its interpretation as a second quantized string partition function, we will  show that $Z_{+}(\rho,\ssigma)$ admits a natural  non-perturbative completion, that we denote by $Z(\rho,\ssigma)$. The free energy $F(\rho, \ssigma) = \log Z(\rho, \ssigma)$ of this extended theory is defined by integrating the CFT partition function against an integration kernel given by the full $\Gamma_{2,2}$ Narain partition sum \eqref{narain0} of the two-torus. We will present evidence that this procedure  is well defined for arbitrary seed CFTs with central charge $c\leq 6$,\footnote{At $c=6$, the convergence of our expressions will turn out to be more delicate than at $c<6$. We will not discuss these subtleties in this paper.} and that the resulting partition function, given in equation \eqref{partfinal}, 
can be interpreted as the grand canonical partition function of an extended $T\bar{T}$-deformed symmetric product CFT.

Moreover, by virtue of the $T$-duality symmetry and spectral characteristics of the Narain partition function, we find that the deformed free energy $F(\rho,\ssigma)$ exhibits a number of remarkable properties listed in the concluding section \ref{sec:conclusions}. Foremost, it is manifestly invariant under a strong weak duality symmetry that inverts the $T\bar{T}$ coupling $\lambda$ to $1/\lambda$. This $\Z_2$ symmetry is part of a large O$(2,2,\Z)$ duality group that includes a novel ${\rm{\rm PSL}}(2,\mathbb{Z})$ S-duality invariance that acts on the modular parameter $\rho$. The emergence of this duality symmetry eliminates the problem that the deformed energies become complex at large coupling and indicates that our generalized $T\bar{T}$-deformation may admit a UV complete description. 
	
This paper is organized as follows. After collecting some preliminary details in section \ref{sec:prelim}, we study the free energy of the grand canonical symmetric product CFT and show that it can be written an integral representation as the one-loop string path-integral in section \ref{sec:partionfunction}. In section \ref{sec:sdualityinv} we introduce and study the partition function of the S-duality invariant extension of the $T\bar{T}$-deformed theory and exhibit its special spectral properties. We summarize our main results in the concluding section \ref{sec:conclusions}. Some technical aspects are deferred to the appendices.

\section{Preliminaries}
\label{sec:prelim}
\vspace{-2mm}
	
We first introduce some preliminaries that will play a key role in our main story: the partition sum of a $T\bar{T}$ deformed CFT, 
the DMVV formula for the partition function of a symmetric product CFT, and the Narain partition function with a $\mathbb{T}^2$ target space. The reader familiar with these topics can skip this section.
	
\vspace{-1mm}
	
\subsection{$T\bar{T}$ deformed CFT partition function}
\vspace{-2mm}
The torus partition function of the $T\bar{T}$ deformed CFT with undeformed energy and momentum spectrum $(E_i, j_i)$ takes the form
\bea
	\label{zonett}
\sum_{i} \exp\bigl({2\pi \i (\ssigma_1 j_i + \i \ssigma_2 \mathcal{E}_i(\lambda) )}\bigr)  
\eea
where $\lambda$ is the $T\bar{T}$ deformation coupling, $\mathcal{E}_i(\lambda)$ is given in \eqref{edeform}, and $\ssigma = \ssigma_1 + \i \ssigma_2$ defines the complex structure modulus of the torus. We define the expression in (\ref{zonett}) as $Z_1(\rho_2,\tau)$ where $\rho_2 \equiv \frac{\tau_2}{\lambda}$. This partition function exhibits invariance under the ${\rm SL}(2,\mathbb{Z})$ modular transformations \cite{Datta:2018thy}
\bea
	Z_1\Bigl(\rho_2, \frac{a \ssigma + b}{c\ssigma + d}\Bigr) \is Z_1(\rho_2,\tau), \quad \quad \rho_2 \equiv \frac{\ssigma_2}{\lambda}
\eea
indicating that it can be obtained by an intrinsically modular invariant path integral computation. The $T\bar{T}$ deformation is the unique universal deformation of individual energy eigenvalues that preserves modular invariance  \cite{Aharony:2018bad}.

Alternatively, we can define the $T \bar{T}$ deformed partition function as the following integral transform of the undeformed CFT partition function
\bea
	\label{gausstrafo} Z_1(\rho_2,\ssigma)  \is \rho_2  \int_{\mathbb{H}_2}
	\frac{d^2\ttau}{\ttau_2^2} \, e^{-\mbox{\footnotesize $\frac{\pi \rho_2}{\ssigma_2 \ttau_2}|\ssigma\!-\! \ttau|^2$}} \, Z_{\rm CFT}(\ttau) \nn\\[-2mm]\\[-2mm]\nn\is 
	\frac{1}{2}\int_{\mathcal{F}} \frac{d^2\ttau}{\ttau_2^2} \hatK_1(\ttau;\rho_2,\ssigma) \, Z_{\rm CFT}(\ttau) \, 
\eea 
where ${\cal F} = \{|\sigma_1| < \half, |\sigma| > 1\}$ is the usual fundamental domain and $\hatK_1$ the modular invariant Poincar\'e series\footnote{Here $\gamma \ttau$ is shorthand for $\gamma \ttau \! =\! \mbox{\small $\frac{a\ttau + b}{c \ttau + d}$}$ with {\scriptsize $\bigl(\begin{matrix}{a}\!\!&\!\! {b}\\[-.75mm] {c}\!\!&\!\!{d}\end{matrix} \bigr)$}$\spc\in\spc{\rm{\rm PSL}}(2,\mathbb{Z})$ and  $(\gamma\ttau)_2$ is a shorthand for $\im(\gamma\ttau)$.} 
\bea
	\label{ksumzero}
	\hatK_1(\ttau;\rho_2,\ssigma) \is 2\rho_2\!\! \sum_{\gamma\in {\rm PSL}(2,\mathbb{Z})} \exp\Bigl({-\frac{\pi \rho_2} {\ssigma_2 (\gamma\ttau)_2}|\ssigma\!-\! \gamma\ttau|^2}\Bigr).
\eea 
This kernel can be identified with the partition function of the seed CFT coupled to a complex scalar field $X(z,\bar{z})$ that maps a dynamical worldsheet with complex structure modulus $\ttau$ into a target space torus with metric
\bea
	\label{torusm}
	G_{ab} dx^a dx^b \is \frac{\rho_2}{\raisebox{2pt}{$\ssigma_2$}}\spc |\spc dx_1+\ssigma\spc dx_2\spc |^2
\eea
and with a target space $B$-field $B_{ab} = B\epsilon_{ab}$ turned on and tuned such that $\sqrt{G}+\i B= 0$.\footnote{Note that this requires imaginary $B$-field.} The expression \eqref{ksumzero} is for unit wrapping (as we review below) and does not include any oscillator contributions, because they cancel with Fadeev-Popov determinants one gets after gauge fixing \cite{Dubovsky:2018bmo,Callebaut:2019omt}. 

The resulting $T\bar{T}$ deformed partition function \eqref{zonett} is real and finite provided that the coupling is restricted to the range $\lambda \leq 6/c$ \cite{Cardy:2022mhn}. 

To derive \eqref{ksumzero}, we first note that	mappings $X: \mathbb{T}^2 \to \mathbb{T}^2$ from a worldsheet torus to a target space torus are labeled by two pairs of winding numbers $w_1 = (m_1, n_1)$ and $w_0 = (m_0,n_0)$. 
The classical solution to the equations of motion $\p \bar \p X_a^{\rm cl} = 0$ with winding numbers ${w} = (w_1,w_0)$ takes the form 
\bea
	\label{xclass}
	X_a^{\rm cl}(z,\bar z) \is \frac{1}{2\i\ttau_{2}{}_{\strut}} \bigl((n_a\! -\nspc m_a \bar \ttau)\, z  \, - \, (n_a \! -\nspc m_a \ttau)\, \bar z\bigr).
\eea
Plugging this classical solution \eqref{xclass}  into the torus sigma model action 
\bea
\label{stringact}
S \is \frac{1}{2\pi} \int\!d^2 z (G_{ab} + B_{ab}) \partial X^a \overline{\partial} X^b.
\eea
with metric \eqref{torusm} and $B_{ab}=\i\rho_2\epsilon_{ab}$ gives
\bea
	\label{sclass}
	S_{\rm cl}(\rho_2,\spc \ssigma,\spc \ttau, \spc w)  \is
	\frac{\pi\rho_2 } {\ttau_2\spc \ssigma_2} | n_1\! +\nspc n_0 \ssigma \nspc -\nspc  m_1 \ttau \! -\nspc  m_0 \ssigma \ttau| ^2 
\eea
The kernel $\hatK_1(\ttau;\rho_2,\ssigma)$ is the classical partition function given by the sum
\bea
	\label{zsum}
	\hatK_1(\ttau;\rho_2,\ssigma) \is 2\rho_2 
	\sum_{w} \; e^{-S_{\rm cl}(\rho_2,\spc \ssigma,\spc \ttau,\spc w)}  
\eea
where $w$ is restricted to the class of maps with wrapping number one: ${\rm gcd}(n_0,m_0) = {\rm gcd}(n_1,m_1) =1$, and $n_0 m_1 - m_0 n_1 = 1.$ The factor of ${1}/{2}$  in the second equality of (\ref{gausstrafo}) can be interpreted as the identification between $w$ and $-w$. We will make this identification in what follows.

\smallskip

\noindent
\subsection{Partition function of a symmetric product CFT}
	
\vspace{-2mm}
	
\noindent
Define the grand canonical partition function of a symmetric product CFT by  
\bea
	\label{sumz}
	Z_+(\rho, \ssigma)  \is 
	1 + \sum_{N=1}^{\infty} \, p^N  Z_N(\ssigma), \qquad \quad p \equiv e^{2\pi \i \rho}  
\eea
where $Z_N(\ssigma)$ denotes the partition function of the $N$-fold symmetric product of some given seed CFT with partition function $Z_1(\ssigma)\coloneqq Z_{\rm CFT}(\ssigma)$. The parameter $p=e^{2\pi \i \rho}$ denotes a complex fugacity that governs the relative weight of symmetric product CFTs with different order $N$.
The total partition function $Z_+(\rho,\ssigma)$ depends on the modular shape parameter $\ssigma$ of the torus and the fugacity parameter $\rho$. 
	
The result that we will obtain for the $T \bar{T}$ deformed partition function of a symmetric product CFT will be a generalization of  the familiar DMVV formula \cite{Dijkgraaf:1996elliptic} for the weighted sum \eqref{sumz} of chiral elliptic genus partition functions 
\begin{align}
	\label{dmvv}
	Z_{{\rm DMVV}}(p,\ssigma) =
	\prod_{d>0}\prod_{m\geq 0}\frac{1}{(1-p^{\spc d}\spc q^m)^{c(md)}\!} 
\end{align}
Here $c(n)$ counts the degeneracy of states with conformal weight $n$ in the seed CFT. The key idea behind the DMVV formula is that the Hilbert space of the $N^{\text{th}}$ symmetric product CFT splits up into a sum over twisted sectors, labeled by conjugacy classes of the permutation group $S_N$. Each twisted sector, in turn, factorizes into a tensor product of long string sectors, labeled by cyclic permutation of order $d$.
	
	The free energy $F_{\rm DMVV} = \log Z_{{\rm DMVV}}$ associated with the chiral symmetric product elliptic genus can be expressed as a sum over positive $N$ of Hecke operators $\tildeT_N$ acting on the chiral seed partition function\footnote{We apologize to the reader for using the same notation $N$ for the order of the Hecke operator as for the order of the symmetric product. The two notions are related but not identical. The $N$ in $\widehat{T}_N$ labels a wrapping number. The path-integral for $Z_N(\ssigma)$ contains multiple wrapping sectors. The total wrapping number of all sectors, as well as the maximal possible wrapping number, is~equal~to~$N$.}
	\bea
	\label{fheck}
	F_{\rm DMVV}(\rho,\ssigma)  \is \sum_{N=1}^\infty \tildeT_N \chi_1(\rho,\ssigma) 
	\eea
Here  $ \chi_1(\rho,\ssigma) \,=\, e^{2\pi \i \rho} \sum_n c(n) e^{2\pi \i n \ssigma}$ denotes the chiral elliptic genus partition function of the seed CFT and the action of the Hecke operator $\tildeT_N$ on a weight-0 modular form $\phi(\rho,\ssigma)$  is defined~as
\bea
	\label{heckeact}
	{\tildeT}_N \phi(\rho,\ssigma) = \frac{1}{N} \sum_{\substack{ad = N,\, d>0 \\ b \,{\rm mod}\, d}} \phi\Bigl(N\rho,\frac{a\ssigma+b}{d}\Bigr).
\eea
$\tildeT_N\phi$ has the geometric interpretation as a modular invariant sum $\tildeT_N \phi = \frac 1 N \sum_f f^* \phi$ over the pullbacks of holomorphic linear maps $f: \mathbb{T}^2 \to \mathbb{T}^2$ of degree $N$ from the torus to itself. This geometric representation points to an interpretation of $F_{\rm DMVV}$ as the one-loop partition function of a second quantized string theory with the two torus as target space.

Via a straightforward generalization of the original derivation of \cite{Dijkgraaf:1996elliptic}, one can show that the grand canonical partition function $Z_+(\rho,\ssigma)$ of the non-chiral symmetric product CFT defined in \eqref{sumz} is given by 
the following formula:
\bea
\label{zzcft}
    Z_+(\rho,\ssigma) 
    \is 
 \prod_{\substack{d>0\\ m\in\mathbb{Z}}}\, \prod_{\substack{i | j_i= md}}\frac{1}{1- p^{\spc d} \spc e^{\mbox{\footnotesize $\frac{2\pi \i}{d}(\ssigma_1j_i+\i \ssigma_2 E_i)$}}}\, .
\eea
The power of $p$ keeps track of the total wrapping number of the string worldsheet. Similarly as the DMVV free energy, the non-chiral free energy $F(\rho,\ssigma) =\log Z(\rho,\ssigma)$ can be written as a sum of Hecke operators acting on the non-chiral seed CFT partition function
\bea
	\label{ffheck}
	F(\rho,\ssigma)  \is \sum_{N=1}^\infty \tildeT_N  \spc Z_1(\rho,\ssigma) 
	\eea
	with  $Z_1(\rho,\ssigma) = e^{2\pi \i \rho} Z_1(\ssigma)$ and with  $\tildeT_N$ defined via \eqref{heckeact}.
	This representation of the free energy will be our starting point for our computation of the $T\Bar{T}$ deformed grand canonical symmetric product partition function.

\subsection{The $\Gamma_{2,2}$ Narain partition sum} 
\label{sec:gamma22narain}

A central player in our story is the Narain partition sum of the gaussian sigma model with a $\mathbb{T}^2$ target space  with metric \eqref{torusm} and general B-field modulus $b = \rho_1$.
\bea
	\label{narain0}
	\hatK_{c=2} 
	(\rho,\ssigma,\ttau) \, = \,  \rho_2\!\!\sum_{\vec{\spc n},\vec{\spc w}\spc\in\spc \mathbb{Z}^2\strut}\!\! e^{\, \mbox{\footnotesize $\frac{\i\pi}{2\ttau_2\ssigma_2}\bigl({\rho}{\spc}|n_2\!+\nspc n_1\ssigma\!-\nspc\ttau(w_1\!+\nspc w_2\ssigma)|^2\! -{\Bar{\rho}\spc}|n_2\!+\nspc n_1\Bar{\ssigma}\! -\ttau (w_1\!+\nspc w_2\Bar{\ssigma})|^2\bigr)^{\strut}$}}\nn\\[-7mm]
\eea 
This $\Gamma_{2,2}$  Narain partition sum, when viewed as a function of the three complex moduli $\ttau$, $\rho$, and $\ssigma$, satisfies a number of remarkable properties.

The total  $\Gamma_{2,2}$ Narain sum \eqref{narain0} is invariant under the extended $T$-duality group
\bea
	{\rm O}(2,2;\mathbb{Z})\simeq {\rm{\rm PSL}}(2,\mathbb{Z})\times {\rm{\rm PSL}}(2,\mathbb{Z})\rtimes \mathbb{Z}^2_2
\eea
given by the product of the modular group acting on the target space modulus $\ssigma$ and the stringy T-duality group acting on complex K\"ahler modulus $\rho$ via
\bea
	\rho \; \to\;  \tilde{\rho} \, = \, \frac{{a} \rho +{b} } {{c} \rho + {d}}, \; \quad \qquad \Bigl(\, \mbox{$\begin{matrix}{a}\!&\! {b}\\[-.5mm] {c}\!&\!{d}\end{matrix}$} \, \Bigr) \in\, {\rm{\rm PSL}}(2,\mathbb{Z}),
\eea
times a $\mathbb{Z}_2$ mirror symmetry that interchanges $\ssigma$ and $\rho$ and a $\mathbb{Z}_2$ symmetry that simultaneous flips the sign of the real part of $\ssigma$ and $\rho$. {Remarkably, it also exhibits a triality symmetry under permutations of the three moduli $\ttau$, $\ssigma$ and $\rho$ \cite{Dijkgraaf:1987jt}.} 

The Narain partition sum $\hatK_{c=2}$ can be written as an infinite sum over terms with fixed torus wrapping number $N$ 
\bea
\label{knarainsum}
	\hatK_{c=2}
	\is  \hatK_{+} + \hatK_{0} + \hatK_{-}, \qquad  \ \ \hatK_{\pm} 
 = \sum_{N>0} \widehat{T}_N \hatK_{\pm 1} 
\eea
where $\hatK_{0}$ and $\hatK_{\pm N} = \widehat{T}_N K_{\pm 1}$ represent the Narain sum restricted to wrapping number $0$ or $\pm N$, respectively.
The zero and $\pm 1$ wrapping number terms admit the following Poincar\'{e} series representation \cite{Benjamin:2021hcft}
\bea
	\hatK_0(\rho,\ssigma,\ttau) \is \rho_2+2\rho_2 \sum_{n=1}^\infty \sum_{\gamma,\tilde\gamma\in\Gamma_{\infty}\!\backslash {\rm{\rm PSL}}(2,\mathbb{Z})}
	e^{-\frac{\pi n^2 \rho_2}{(\gamma\ssigma)_2 (\tilde\gamma\ttau)_2}},\\[2mm]
	\label{kpm1}
	\hatK_{\pm 1}(\rho,\ssigma,\ttau)= & & \hspace{-6mm} 2\rho_2 \!\!\! 
	\sum_{\gamma\in {\rm{\rm PSL}}(2,\mathbb{Z})}\!\! e^{\mbox{\footnotesize $\frac{\i\pi} 2\bigl(\frac{\rho}{\ssigma_2 (\gamma\ttau)_2}|\ssigma\mp \gamma\ttau|^2 - \frac{\bar\rho}{\ssigma_2(\gamma\ttau)_2}|\ssigma\mp\gamma\bar\ttau|^2\bigr)$}}.
\eea 
Note that if we set $\bar\rho = 0$ and $\rho = 2\i\rho_2$, the wrapping number 1 term $\hatK_{+1}$ coincides with the integration kernel \eqref{ksumzero} used to define the $T\bar{T}$ deformed partition function.
The zero wrapping term $\hatK_0$ coincides with the trace of the heat kernel defined on the torus target space.

\section{Partition function of $T\bar{T}$ deformed symmetric product CFT}
\label{sec:partionfunction}
	
\vspace{-1mm}

\noindent
We now turn to study the grand canonical partition function $Z_{+}(\rho,\ssigma)$ of the $T\bar{T}$-deformed symmetric product CFT.  First, we compute $Z_{+}(\rho,\ssigma)$ via a combinatoric argument and by applying the Smirnov-Zamolodchikov formula \eqref{edeform} for the deformed energy spectrum to the twisted sectors of the orbifold CFT. We then introduce an unoriented generalization of the symmetric product CFT and give another derivation of the free energy using its representation \eqref{ffheck} as a sum over a string worldsheets  with non-zero wrapping numbers. 
Finally, we will rewrite the free energy in terms of a single integral kernel applied to the seed CFT partition function. In the next section we will use the link between this integral kernel and the $\Gamma_{2,2}$ Narain sum $\hatK$ to define an $S$-duality invariant extension of the grand canonical partition function.

\subsection{$T\bar{T}$ deformed symmetric product CFT}
\vspace{-1mm}

The grand canonical partition function \eqref{symsum}
can be defined as a trace over the Hilbert space of an infinite direct sum of $N$-fold symmetric product CFTs 
\bea
{\cal H}_+ \is \bigoplus_{N>0} {\cal H}_N \qquad \quad 
{\cal H}_N \equiv  {\cal H}(S^N{\rm CFT})
\eea
Let  $\hat{N}$ denote the operator that counts the order of the symmetric product, $\hat{J}$ the momentum operator, and $\hat{H}(\lambda)$ the deformed Hamiltonian.
Equation \eqref{symsum} can then be written as a trace 
 \bea
\hatZ_+(\rho,\ssigma)  \is 1 + \sum_{N>0} \, {p}^N  Z_N(\rho_2, \ssigma) = \tr_{{\cal H}_+}\bigl(e^{2\pi \i \hat{N}\rho} e^{ 2\pi \i (\ssigma_1 \hat{J} + \i\ssigma_2 \hat{H}{(\lambda)})}\bigr), ~~~~\rho_2 = \frac{\tau_2}{\lambda}
\eea
This partition function depends on the modular parameter $\ssigma$ of the torus and a complex coupling  $\rho = \rho_1 + \i \rho_2$  that encodes the $T\bar{T}$ coupling $\lambda$ and fugacity parameter that keeps track of $N$.  Below we will derive the following result
\bea
\label{zttt}
\hatZ_+(\rho,\ssigma)  \is  \prod_{\substack{d>0\\ m\in\mathbb{Z}}}\spc \prod_{\substack{i | j_i=md}} \, \frac{1}{1-p^{\spc d}\spc e^{\frac{2\pi \i}{d} (\ssigma_1 j_i+\i\ssigma_2 \mathcal{E}_i({\lambda}/{d^2}))}}	
\eea
with $p=e^{2\pi \i \rho}$ and ${\cal E}(\lambda)$ the deformed energy given in \eqref{edeform}.

Formula \eqref{zttt} takes the expected form, as it arises simply by deforming the energy levels of the corresponding CFT partition function \eqref{zzcft}. The only aspect that needs some explanation is the adaptive rescaling $\lambda \to  \lambda/d^2$ of the $T\bar{T}$ coupling in the long string sectors with integer length $d$. 
We claim that this rescaling among different winding sectors is necessary to ensure that the {dimensionful} $T\bar{T}$ coupling is the same across all long string sectors.  To see this, recall that we have chosen units so that the space dimension of the deformed CFT is a circle with unit radius $R=1$. All quantities, including~$\lambda$, are made dimensionless by multiplying by the appropriate power of the circle radius $R$. Let $\bar{\lambda}$ denote the dimensionful $T\bar{T}$ coupling. Since the $T\bar{T}$ operator has mass dimension $4$, $\bar\lambda$ has mass dimension $-2$, or length dimension $2$. In the unit winding sector, this means that the two coupling are related via $\lambda = \bar\lambda/R^2$.

Let us briefly recall how the long string phenomenon comes about \cite{Dijkgraaf:1996elliptic}. The $S_N$ symmetry is a gauge symmetry of the symmetric orbifold. We can thus define twisted sectors labeled by conjugacy classes of the orbifold group $S_N$ 
\be
\cH(S^N{\rm CFT}) = \bigoplus_{\{{N_d}\}} \cH_{\{{N_d}\}}.
\ee
where we used that the conjugacy classes $[g]$ of $S_N$ are characterized
by partitions $\{N_d\}$ of $N$ with $\sum_{d} d N_d = N$. Here $N_d$ denotes the multiplicity of the cyclic permutation $(d)$
of $d$ elements in the decomposition of $[g] = (1)^{N_1}(2)^{N_2} \ldots (s)^{N_s}$. 
In each twisted sector, one needs to impose invariance under the centralizer subgroup $C_g = 
\prod_{d=1}^s \, S_{N_d} \times \Z^{N_d}_d,$
where each $S_{N_d}$ permutes the $N_d$ cycles $(d)$, while each
$\Z_d$ acts within one particular cycle $(d)$. Correspondingly,  we can decompose each
twisted sector as
\be
\label{hdeco}
{\cH}_{\{N_d\}} = \bigotimes_{d>0} \, S^{N_d} \cH_{(d)} \qquad \quad
S^N\cH = \Bigl(\underbrace{\cH \otimes \ldots \otimes \cH}_{N\ \rm times}
\Bigr)^{S_N}.
\ee 
The spaces $\cH_{(d)}$ 
in this decomposition denote
the $\Z_d$ invariant subsector of the space of states with winding number $d$. In this twisted sector, the momentum per winding can be fractional of the form $j^{(n)}_i/d$ with $j_i^{(n)} \in \Z$. The $\Z_d$-invariant subspace consists of those states 
for which these fractional momenta combined add up to an integer. 
 
 From the above description, we see that the long strings wind around $d$ times and thus have spatial length $d R$. Their energy and momentum levels are thus reduced by a factor $d$. Moreover, the rescaling of the spatial length means that, relative to the total length $R d$ of the long string, the dimensionless $T\bar{T}$ coupling in the long string sector is $\lambda_d = \bar\lambda/(R d)^2 = \lambda/d^2$. Combined these observations lead to the announced result \eqref{zttt} for the deformed grand canonical partition function. As we will see in section \ref{subsec:freeEnergy}, it can also be derived by integrating the partition function of the seed CFT against the deformation kernel $\mathcal{K}_+$ defined in equation (\ref{knarainsum})
 \begin{equation}
    \log Z_+(\rho,\tau) = {1\over 2}\int_{\mathcal{F}}{d^2\sigma\over\sigma_2^2}\mathcal{K}_+(\rho,\tau,\sigma)Z_{\rm CFT}(\sigma).
 \end{equation}
 Note that the long string sectors with one definite sign of wrapping number contributes. Hence we can think of $Z_+(\rho,\ssigma)$ as the partition function of an oriented second quantized string theory with the same world-volume theory as the $T\bar{T}$-deformed seed CFT. This interpretation of $Z_+(\rho,\ssigma)$  reflects the combinatoric equivalence between the second quantized Hilbert space and the direct sum over all symmetric products of the single particle Hilbert space.

\subsection{Unoriented $T\bar{T}$ deformed symmetric product CFT}
\vspace{-1mm}

We now make the logical next step of including the contribution of all wrapping sectors, including sectors with negative and zero wrapping numbers.  The negative wrapping sectors are naturally  interpreted as the mirror image of the positive wrapping sectors. 

Specifically, we wish to study the partition function 
of the extended $T\bar{T}$-deformed symmetric product CFT defined via the trace
 \bea
 \label{ztotal}
Z(\rho,\ssigma)  \is \tr_{{\cal H}}\bigl(e^{2\pi \i \hat{N}\rho} e^{ 2\pi \i (\ssigma_1 \hat{J} + \i\ssigma_2 \hat{H}{(\lambda)})}\bigr) 
\eea
where ${\cal H}$ denotes the total Hilbert space given by the tensor product of three sectors
\bea
\label{htensor}
{\cal H} \is {\cal H}_+ \otimes {\cal H}_0 \otimes {\cal H}_-
\eea
Here ${\cal H}_\pm$ are the positive and negative wrapping sectors. They are given by the infinite direct sum ${\cal H}_\pm = \bigoplus_{N>0} {\cal H}_{\pm N}$
with ${\cal H}_N$ and  ${\cal H}_{-N}$ the $N$-fold symmetric product Hilbert space of the seed CFT and of the orientation reversed seed CFT, respectively. Orientation reversal is defined by flipping the sign of all momenta $j_i$. So the  sectors ${\cal H}_N$ and ${\cal H}_{-N}$ look identical in the case that the spectrum of the seed CFT is parity-symmetric. They still contribute separately and differently to the partition function, because of the presence of the chemical potential. 
 The zero-wrapping sector ${\cal H}_0$ has a less obvious CFT interpretation. We will study this sector in more detail in section \ref{sec:sdualityinv}.
 
The tensor product Hilbert space \eqref{htensor} includes long string wrapping sectors of both signs. It thus represents an unoriented second quantized string theory.  The partition function \eqref{ztotal} of this unoriented theory factorizes into a product of three factors
\bea
\label{ztot}
Z(\rho,\ssigma) \is  Z_+(\rho,\ssigma)  Z_0(\rho,\ssigma) Z_-(\rho,\ssigma)  
\label{zhat}
\eea
each given by the trace over the corresponding Hilbert space subfactor.

Multiplying the result \eqref{ztotal} with its orientation reversed copy and plugging in the explicit form \eqref{edeform} of the deformed energy gives that
\bea
\label{zpm}
Z_+(\rho,\ssigma) Z_-(\rho,\ssigma)  \is \prod_{\substack{d\neq 0 \\ m\in\mathbb{Z}}}\spc \prod_{\substack{i| j_i=md}} \, 
\frac{1^{{}_{\strut}}}{1- p_1^{\spc d}\spc q_1^{\spc m} \spc e\raisebox{4pt}{\footnotesize $-2\pi\ssigma_2  \sqrt{ (d/\lambda)^2\! +\nspc m^2\! +\nspc 2E_i/ \lambda}$\, }}
\eea
with $p_1=e^{2\pi \i \rho_1}$ and $q_1=e^{2\pi \i \ssigma_1}$.
The above product includes long strings with all wrapping numbers $d$ except zero. It is therefore natural to consider the extended symmetric product partition function \eqref{zhat} that includes the factor with $d$ set to zero
\bea
\label{zzero}
Z_0(\rho,\ssigma) \is e^{-\rho_2 V_0(\lambda) } \prod_{i \in {\mathcal S}} \prod_{m \in \Z}\, 
\frac{1^{{}_{\strut}}}{1- q_1^m\spc e\raisebox{6pt}{\footnotesize $ -2\pi\ssigma_2 \sqrt{\nspc m^2\! +\nspc 2E_i/ \lambda}$}},
\eea
where ${\cal S}$ denotes the set of all spin zero states in the seed CFT. Here we included a possible vacuum energy contribution $V_0(\lambda)$. 
This zero wrapping number partition function may look a bit mysterious from the CFT perspective but has a clear string theoretic interpretation: it represents the partition function of an infinite tower of free spinless particles moving on the torus with metric \eqref{torusm}, with each particle corresponding to a $j_i=0$ state in the seed CFT. We will study this zero-wrapping sector in more detail in section \ref{sec:sdualityinv}, where we will derive an explicit formula for the vacuum energy $V_0(\lambda)$ based on duality symmetry.

\subsection{$T\bar{T}$ deformed grand canonical free energy}\label{subsec:freeEnergy}
\vspace{-1mm}

A second instructive derivation of the partition functions \eqref{zttt} and \eqref{zpm} makes use of the fact that the corresponding free energy $F_\pm(\rho,\ssigma) =\log Z_\pm(\rho,\ssigma)$ can be written as a sum over positive integers $N$ 
 of Hecke operators $\tildeT_N$ acting on the seed $T\bar{T}$-deformed seed CFT partition function
\bea
\label{ffheckt}
\hatF_\pm (\rho,\ssigma)  \is  \sum_{N=1}^\infty \tildeT_N  \spc Z_{\pm 1}(\rho,\ssigma) \\[2mm]
Z_{\pm1}(\rho,\ssigma)
= & & \hspace{-6mm} p_\mp  \sum_{i} e^{\mbox{\footnotesize $\pm 2\pi \i j_i \ssigma_1 - 2\pi \ssigma_2 \mathcal{E}_i(\lambda)$}}  
\label{zexplicit}
	\eea 
with $p_\pm = e^{2\pi \i (\pm \rho_1 + \i\rho_2)}$. Here the action of the Hecke operators $\tildeT_N$ is defined via
\bea
\label{heckt}
    \tildeT_N Z_{\pm 1}(\rho,\ssigma) \is  \frac{1}{N} \sum_{\substack{ad = N,\, d>0 \\ b \,{\rm mod}\, d}} Z_{\pm 1}\Bigl(N\rho, \frac{a\ssigma+b}{d}\Bigr).
\eea 
The formula \eqref{ffheckt} is the direct generalization of the formula \eqref{ffheck} for the free energy $F(\rho_1,\sigma)$ of the non-chiral symmetric product CFT.
To evaluate this expression, we first plug in the explicit form \eqref{zexplicit} of the deformed seed partition sum and then perform the summation over $b$. This gives the periodic delta function
\bea
\frac 1 N  \sum_{\substack{ b \,{\rm mod}\, d}} e^{\mbox{\footnotesize $\frac{2\pi \i}{d} j_i (a\ssigma_1 + b)$}} \is   \frac 1 a \, e^{\mbox{\footnotesize $\frac{2\pi \i}{d}a j_i\ssigma_1$}} \sum_{m\in \mathbb{Z}} \delta_{j_i, md}, \qquad \quad a = \frac N d
\eea
restricting the value of the momentum $j_i$ to be an integer multiple of $d$. This indicates that $d$ plays the physical role of the winding number of the mapping from the worldsheet torus into the target space torus.
The sum over $N$ can now be evaluated as follows: 
\bea
   \hatF_\pm(\rho,\ssigma) \is    \sum_{N>0} \, \sum_{\substack{ad = N\\[.5mm]d>0}}  \, \sum_{m\in\mathbb{Z}} \sum_{i\, | \, j_i = m d}\,  \, \frac{1}{a}\spc p_\mp^{N} e^{\mbox{\footnotesize $\frac{2\pi \i}{d} a (\pm j_i \ssigma_1 + \i\ssigma_2 \mathcal{E}_i(\lambda/d^2))$}},\\[1mm]\is  \;  \sum_{\substack{a,d>0\\m\in\mathbb{Z}}} \; \sum_{i\, | \, j_i = m d} \, \; \frac{1}{a}\spc p_\mp^{ad}  e^{\mbox{\footnotesize $\frac{2\pi \i}{d} a (\pm j_i \ssigma_1 + \i\ssigma_2 \mathcal{E}_i(\lambda/d^2))$}},\\[1mm]
    \is \; \sum_{\substack{d>0\\m\in\mathbb{Z}}} \; \sum_{i\, | \, j_i = m d} - \log \bigl(1- p_\mp^{d} e^{\mbox{\footnotesize $\pm 2\pi \i m \ssigma_1 - {2\pi}\ssigma_2 \frac 1 d \mathcal{E}_i(\lambda/d^2)$}}\bigr). \ \
    \label{logz}
    \eea
The rescaling of the $T \bar{T}$  coupling to $\lambda/d^2$ follows from the relation $\lambda = \ssigma_2/\rho_2$ and the fact that the Hecke operator \eqref{heckt} acts by replacing $(\rho_2,\ssigma_2)$ by $(N \rho_2, a\ssigma_2/d)$ with $N=ad$. Taking the exponent of \eqref{logz} leads to the final result \eqref{zttt} and \eqref{zpm}  for the partition function at non-zero wrapping.

The  free energy $\hatF_\pm(\rho,\ssigma)$ of the deformed symmetric product CFT can be expressed as the integral  over the fundamental domain of the undeformed partition function $Z_{\rm CFT}(\ttau)$ of the seed CFT times a diffusion kernel $\hatK_{\pm}$ 
\bea
	\label{string}
	\hatF_\pm(\rho,\ssigma) \is \frac{1}{2}\int^{\strut}_{\mathcal{F}\strut}\frac{d^{2}\ttau{}}{\ttau_{2}^2} \, \hatK_\pm( \rho,\ssigma,\ttau)\, Z_{\rm CFT}(\ttau)\  \, 
\eea 
given by the sum over all positive integers $N$ of Hecke operators \eqref{heckeact}
\bea
	\label{kpm}
\hatK_\pm (\rho, \ssigma,\ttau)^{\strut}
	\is \sum_{N> 0} \tildeT_{N} \hatK_{\pm 1} (\rho,\ssigma,\ttau)	
\eea
acting on the modular invariant diffusion kernels in the $N\!=\! 1$ wrapping~sector. The positive and negative wrapping number contributions $\hatK_N$ and $\hatK_{-N}$ are related via $\hatK_{-N}(\rho, \tau, \sigma) = \hatK_N(-\bar{\rho}, -\bar{\tau}, \sigma)$,  i.e. an orientation reversing involution that interchanges the left-moving and right-moving sector of the CFT. 
 Equations \eqref{string}-\eqref{kpm} combine and generalize the formulas \eqref{gausstrafo} and \eqref{fheck} for the deformed  partition function and the symmetric product free energy, respectively.  The formula \eqref{string} yields a finite real result as long as $c \leq 6$ and $\lambda\leq 6/c$.

\noindent
\section{S-duality invariant $T\bar{T}$ deformed CFT partition function}
\label{sec:sdualityinv}

Equation \eqref{zpm} has the structure of the partition function of a free second quantized (non-critical) string theory with worldsheet theory 
given by the product of the seed CFT times a free boson sigma model \eqref{stringact} with a $\mathbb{T}^2$ target space  with metric \eqref{torusm} and B-field modulus $b = \rho_1$. This correspondence motivates a natural definition of the $T\bar{T}$ deformed symmetric product CFT that takes this relationship with second quantized (non-critical) string theory seriously. We will argue that this string theory description defines a non-perturbative completion of the $T\bar{T}$ deformed CFT, in the sense that it will allow us to define the deformed partition function for all (non-negative) values of the $T\bar{T}$ coupling.

The diffusion kernels $\hatK_{\pm}(\rho, \ssigma,\ttau)$ defined in \eqref{kpm} have a natural extension obtained by including the zero wrapping number contribution 
\bea
\label{ksum}
\hatK(\rho, \ssigma,\ttau) 
\is \hatK_{+}(\rho, \ssigma,\ttau) + \hatK_0(\rho, \ssigma,\ttau) + \hatK_-(\rho, \ssigma,\ttau).\, 
\eea
Comparing with \eqref{knarainsum} suggests that we should equate this extended kernel $\hatK$ with the $\Gamma_{2,2}$ Narain partition sum $\hatK_{c=2}$ introduced in equation (\ref{narain0}) in section \ref{sec:gamma22narain}. Indeed, the full sum \eqref{narain0} includes the instanton contributions of string worldsheets with arbitrary wrapping numbers. We thus are led to consider the integral transform of the seed CFT partition function 
\bea
	\label{fhat}
	\hatF(\rho,\ssigma) \is  \frac{1}{2}\int_{\mathcal{F}}\frac{d^2\ttau}{\ttau_2^2} \, \hatK(\rho,\ssigma,\ttau)\, Z_{\rm CFT}(\ttau) 
\eea
where $\hatK(\rho,\ssigma,\ttau)$ is given by the full $\Gamma_{2.2}$ Narain sum   \eqref{knarainsum} .
As we will argue in the following, this integral transform yields a unique and finite answer provided that the seed CFT has central charge $c\leq 6$. By construction, the free energy $F(\rho,\ssigma)$ defined by \eqref{fhat} exhibits ${\rm O}(2,2;\mathbb{Z})$  duality  symmetry. This duality group includes a $\mathbb{Z}_2$ mirror map that interchanges $\ssigma$ and $\rho$. Since the $T\bar{T}$ coupling is given by $\lambda ={\ssigma_2}/{\rho_2}$, the mirror map acts via $\lambda \leftrightarrow 1/\lambda$ and thus interchanges strong and weak coupling. We will see that this mirror symmetry is sufficient to remedy the seemingly pathological occurrence of complex energy levels in the strong coupling regime of the $T\bar{T}$ deformed theory.

\subsection{Free energy at zero wrapping number}
	
The free energy at zero wrapping number is formally defined via the integral expression
\bea
\hatF_0(\rho,\ssigma) \is \frac{1}{2}\int^{\strut}_{\mathcal{F}\strut}\frac{d^{2}\ttau{}}{\ttau_{2}^2} \, \hatK_0(\rho,\ssigma,\ttau)\, Z_{\rm CFT}(\ttau)\\[2mm]
\hatK_0(\rho,\ssigma,\ttau) \is \rho_2+2\rho_2 \sum_{n=1}^\infty \sum_{\gamma,\tilde\gamma\in\Gamma_{\infty}\!\backslash {\rm{\rm PSL}}(2,\mathbb{Z})}
	e^{-\frac{\pi n^2 \rho_2}{(\gamma\ssigma)_2 (\tilde\gamma\ttau)_2}}
\eea
The first term $\rho_2$ in the diffusion kernel $\hatK_0$ is the zero winding contribution. Its presence makes the integral divergent for any compact CFT seed with positive central charge $c>0$. However, as we will argue below, this part of the integral can be regularized to give a unique finite answer via analytic continuation.

The integral against the second term of $\hatK_0$ can be evaluated via a standard unfolding trick. The Poincar\'e sum over $\tilde{\gamma} \in$ PSL$(2,\Z)$ can be replaced by an unfolded $\ttau$ integral over $\mathbb{H}_2/\Gamma_{\infty}$, i.e. the infinite strip between $-\frac{1}{2} < \ttau_1 < \frac{1}{2}$.  This allows us to perform the integral over $\ttau_1$. The remaining integral reads as follows
\bea
	\label{fzerozero}
	F_0(\rho,\ssigma) 
	\is  \mu_0 \rho_2 + \frac{\rho_2}{2}\sum_{i\in \mathcal{S}}   \sum_{(m,n)\neq (0,0) }\int_{0}^{\infty} \frac{dy}{y^2}  e^{-\pi \rho_2 \frac{|n+m\ssigma|^2}{y \ssigma_2} -2\pi y E_i}.
\eea
where  we replaced the Poincar\'{e} series by a summation over pairs of integers $m,n$, and $\{ E_i, i\in {\cal S}\}$ denotes the spectrum of zero spin states.  The constant
 $\mu_0$ denotes the divergent integral of the seed CFT partition function over the fundamental domain
\bea
\label{vacenergy}
	   \mu_0  \coloneqq  \frac{1}{2}\int_{\mathcal{F}} \frac{d^2\ttau}{\ttau_2^2} Z_{\rm CFT}(\ttau).
\eea 

To exhibit the physical meaning of \eqref{fzerozero}, we will rewrite it in three different ways. First we use the relationship between the integration kernel ${\cal K}_0$ and the heat kernel on the two-torus to formally equate $F_0$ to a sum over all spin zero states of the logarithm of functional determinants. Second, we explicitly perform the $y$-integral in equation \eqref{fzerozero} to write $F_0$ as a convergent sum over Bessel functions. Third, by first performing a Poisson resummation over $n$ before integrating over $y$, we express $F_0$ as the free energy of an infinite set of particles with mass squared $E_i$.

\def\half{ \mbox{\large ${\frac 1 2}$} }
\begin{enumerate}
\item{Equation \eqref{fzerozero} can be recognized as a sum of logarithms of functional determinants
\bea
\label{fzeroone}
F_0(\rho,\ssigma)\is A \rho_2 -\frac{1}{2}\sum_{i \in \mathcal{S}} \log \det (-\Delta+2E_i) 
\eea
where $\Delta$ denotes the laplacian operator defined on the two-torus with metric \eqref{torusm}. As reviewed in Appendix \ref{sec:heatkernel}, the log of the functional determinant on the torus has the following familiar expression 
in terms of the trace of the heat kernel
\bea
\label{detheat}
	-\log \det\bigl( -\Delta + M^2\bigr) 	\is 
 (2\pi)^2\rho_2 \int_0^\infty \frac{dt}{t} \, K_0(t,x,x)\, e^{-M^2t}.
\eea
The heat kernel on the torus \eqref{torusm} has the following explicit form
\bea
	\label{heatkernel}
	K_{0}(t,x,x) =\frac{1}{4\pi t}\Bigl( 1 +\!\!\!\!\sum_{(n,m)\neq (0,0)}\!\!\! 
	e^{-\pi \rho_2 \frac{|n+m\ssigma|^2}{t \ssigma_2}}\Bigr)
\eea
Comparing equations \eqref{detheat} and \eqref{heatkernel} with the integral expression \eqref{fzerozero} confirms that $F_0(\rho,\ssigma)$ can be rewritten as in \eqref{fzeroone}, provided the finite vacuum energy $\mu_0$ and the divergent constant $A$ are related via 
\bea
\label{adeff}
A = \mu_0 + \sum_{i\in {\cal S}} \int_{-\infty}^\infty\!\! dk \, \pi \sqrt{k^2+ 2E_i}
\eea
The second term  is designed to cancel the divergent contribution to the integral \eqref{detheat} from the  $\frac{1}{4\pi t}$ term. The formula \eqref{fzeroone} makes explicit that the zero wrapping sector describes an infinite set of spinless particles, labeled by the spin zero CFT states, with mass squared equal to $\{2E_i, i \in {\cal S}\}$.}

\medskip

\item{Another source of potential divergence is that some of the energy levels $E_i$ are negative. The particles associated with these states are tachyonic and the corresponding integral over $y$ in \eqref{fzerozero} diverges at the cusp. We can regulate this divergence in the standard way by first assuming that $E_i$ is a complex number with $\text{Re}(E_i)>0$, perform the integral over $y$ and then analytically continue to the physical value of $E_i$. This yields the following expression
	\bea
	\label{fzerotwo}
		F_0(\rho,\ssigma)\is  \mu_0 \rho_2 +  	\sum_{i\in \mathcal{S}}\sum_{(m,n)\neq(0,0)}\sqrt{\frac{2\rho_2\ssigma_2 E_i}{|m\ssigma+n|^2}}K_1\(\sqrt{\frac{8\pi^2\rho_2|m\ssigma+n|^2E_i}{\ssigma_2}}\) \qquad
  \label{eq:sumlogdet}
	\eea
The sum over scalars in (\ref{eq:sumlogdet}) converges when $c\leq 6$ and $ \lambda \leq 6/c$. To see this, we note that the Cardy formula for the scalar density of states grows as $e^{\sqrt{4\pi^2 cE/3}}$ \cite{Cardy:1986ie}, whereas the Bessel function falls off as $e^{-\sqrt{{8\pi^2|m\ssigma+n|^2E}/\lambda}}$. If we choose $\ssigma$ to lie in the standard fundamental domain then $|m \ssigma + n|^2 \geq 1$. Hence the Bessel function falls off fast enough to overcome the Cardy growth precisely in the same regime where the $T\bar{T}$ deformed CFT is well defined. The fact that the two bounds are identical is not a coincidence: they related via a modular $S$ transformation.}

\medskip

\def\sN{\mbox{\small$N$\nspc}}
\item{The identity \eqref{fzeroone} tells us that $Z_0(\rho,\ssigma)$ can be written in the form of a quantum mechanical partition function 
\bea
\label{zzeropsum}
Z_0(\rho,\ssigma)  = e^{F_0(\rho,\ssigma)} \is \tr_{{\cal H}_0} \Bigl(e^{ 2\pi \i (\ssigma_1 \hat{J} + \i\ssigma_2 \hat{H}{(\lambda)})}\Bigr) 
\eea
where ${\cal H}_0$ denotes the second quantized Hilbert space of an infinite set of spinless particles of mass squared $m_i^2\! = 2E_i$ defined on the cylinder with radius $\sqrt{{\rho_2}/{\ssigma_2}}$. Equivalently, we can choose units so that the cylinder radius is $1$ and the scalar particles have mass squared $2E_i/\lambda$ with $\lambda = \ssigma_2/\rho_2$. The explicit form of the Hilbert space can be derived by deconstructing the path-integral representation of the functional determinant. 

An alternative derivation is presented in Appendix \ref{sec:funcdet}, where it is shown that \eqref{fzeroone} can be re-expressed as
\bea
\label{fzerothree}
F_0(\rho,\ssigma)\nspc \is\nspc -\rho_2 V_0(\lambda)
 -  \sum_{i\in {\cal S}} 
\sum_{n \in \Z}  \log\bigl(\nspc 1\! -\nspc e^{2\pi \i \ssigma_1 n - 2\pi \ssigma_2
{\cal E}_{i}(n,\lambda)}\nspc\bigr)\\[-8mm]\nn
\eea
with\\[-8mm]
\bea
\label{vlambda}
V_0(\lambda) \is -A + \sum_{i\in {\cal S}} \sum_{n\in\mathbb{Z}}  \pi \lambda \, {\cal E}_{i}(n,\lambda), \quad \quad
{\cal E}_{i}(n,\lambda) =  {\sqrt{n^2\! +{2E_i}/\lambda }}
\eea
 the energy level of the $n$-th momentum mode of $i$-th complex scalar field, and $A$ the divergent integral given in \eqref{adeff}.
 Equation \eqref{fzerothree} exponentiates to the form anticipated by (\ref{zzero})
\bea
Z_0(\rho,\ssigma)  \is e^{- \rho_2 V_0(\lambda)}  \prod_{i\in {\cal S}} \,\prod_{n\in \mathbb Z} \frac{1}{1- e^{2\pi \i \ssigma_1 n - 2\pi \ssigma_2
{\cal E}_{i}(n,\lambda)}}\, .
\eea
This matches the partition function of a  Fock space labeled by occupation numbers $\sN_{n_i}$ and with the following energy eigenspectrum
\bea
\hat{H}(\lambda) |\{\sN_{n_i}\} \rangle \is \sum_i \sN_{n_i} {\cal E}_i(n_i,\lambda)|\{\sN_{n_i}\}\rangle\\[-6mm]\nn
\eea
plus a constant overall vacuum energy shift equal to $E_{\rm vac} = V_0(\lambda)/\lambda$.
}
\end{enumerate}

     \medskip
     
\subsection{Vacuum energy contribution}

Let $Z_{c_0}(\ttau)$ be the seed CFT partition function with positive central charge $0<c_0\leq 6$.
We wish to compute the vacuum energy contribution $\mu_0$ given by the integral of $Z_{c_0}(\ttau)$ over fundamental domain.
This integral is divergent due to the exponential growth $Z_{c_0}(\ttau) \sim e^{\frac{\pi c_0}{12} \ttau_2}$ at the cusp. We will regulate this integral via an analytic continuation procedure introduced in \cite{Benjamin:2022pnx} in the context of Narain CFTs. Here we generalize the same procedure to arbitrary CFTs with central charge in the specified range.

The key idea is to first multiply the seed partition function $Z_{c_0}(\ttau)$ by some auxiliary modular invariant partition function $\Phi(\ttau)^{d}$ with negative central charge $d =  c - c_0$, as defined via its growth at the cusp. It is natural to choose $\Phi(\ttau)$ to be equal to the inverse of a rational $c=1$ CFT partition function.  The modified seed partition function 
\bea
\label{newzc}
{Z}_c(\ttau) \is \Phi(\ttau)^{c-c_0} Z_{c_0}(\ttau)
\eea 
has effective central charge ${c} = c_0+ d <0$. This removes the exponential growth at the cusp and ensures that the effective energies in the mode expansion all satisfy $E_i\geq 0$. We then perform the integral of ${Z}_c(\ttau)$ over the fundamental domain 
\bea
\label{muzeroint}
\mu_c \is \frac{1}{2}\int_{\mathcal{F}} \frac{d^2\ttau}{\ttau_2^2} Z_{c}(\ttau),
\eea
and obtain a finite result.  We will assume that $\mu_c$ is an analytic function of the effective central charge $c$ of the modified partition function. We can then use analytic continuation to define the value $\mu_{c_0}$ of the integral at the physical value of the central charge. Equating $\mu_0 = \mu_{c_0}$, we thus obtain a regularized version of the original integral. The details of this procedure are outlined in Appendix \ref{sec:integralregapp}.

In Appendix \ref{sec:integralregapp} we generalize the ideas developed in \cite{Benjamin:2022pnx} to derive the following convergent expression for any CFT with central charge $c\leq6$ 
\bea
	\label{finitemu}
	\mu_0  \is  
	\sum_{i\in\mathcal{S}} \, \frac{2\sqrt{q}}{q-1} \, \sqrt{2E_i} \sum_{m=1}^\infty  \frac1m \Bigl(K_1\bigl(2\pi m\sqrt{2E_i/q}\bigr) - K_1\bigl(2\pi m\sqrt{2qE_i}\bigr)\Bigr).
\eea
Here $q$ is some arbitrary real number between $c/6$ and $6/c$. 
Remarkably, if the set of energies $\{E_i\}$ specify the spectrum of spinless states of a modular invariant CFT, then the numerical value on right-hand side is independent of the value of $q$. We checked this statement to high numerical accuracy. It is natural to use this freedom and take the limit as $q$ approaches $1$.
A good argument for taking this limit is that at large energy, the leading term in the sum over Bessel function decays as $e^{-2\sqrt 2 \pi \sqrt{E_i} \sqrt{\text{min}(q, q^{-1}})}$. Hence we maximize the decay by taking $q\to 1$. Comparing this decay with the Cardy growth $e^{2\pi \sqrt{{c E_i}/3}}$ of the scalar energy spectrum \cite{Cardy:1986ie}, we find that (\ref{finitemu}) converges if  $c \leq  6~ \text{min}(q, q^{-1})$. Setting $q=1$ gives the largest range, $c\leq 6$.

An alternative formula for $\mu_0$ that gives a bit more insight into the physical meaning of the vacuum energy contribution is
\bea
\label{munew}
	\mu_0 \is \, \frac{1}{q\nspc - \nspc 1\!} \; \bigl(\spc  \nu_0(q) - q\spc \nu_0(1/q)\bigr)
 \label{eq:alternatemuo}\\[-9mm] \nn
 \eea
 with\\[-8mm]
 \bea
 \label{nuzero}
\nu_0(q) \is 2 \sum_{i\in {\cal S}}\!\! \int\limits_{{\;\;\;\mbox{\scriptsize $
\sqrt{2^{}\ \;\;\;\;\;\,}\!\!\!\!\!\!\!\!\!\!\!\! E_i/\nspc q$}}
}^{\ \;\; \infty}\!\!\! dp \; {2\pi}q\,  \frac{\sqrt{p^2\nspc - 2E_i^{\!}/ q}}{e^{2\pi p} -1} 
 \eea
We (numerically) verified that this expression is well-defined and finite, and equivalent to formula \eqref{finitemu}, over the parameter regime $c \leq 6\min(q,q^{-1})$. 
 We will now present an independent derivation of the result \eqref{eq:alternatemuo}-\eqref{nuzero} based on duality symmetry.

In equation \eqref{vlambda} we found that the total vacuum energy contribution $V_0(\lambda)$ introduced in \eqref{zzero} is given by 
\bea
 V_0(\lambda) \is - \mu_0  +  \sum_{i\in S} \biggl[\, \sum_{n\in\mathbb{Z}}\pi \lambda  \sqrt{n^2\! + \nspc {2E_i}/\nspc{\lambda}}   -  \int_{-\infty}^\infty\!\! dk \, \pi \sqrt{k^2\!+\nspc 2E_i}\, \biggr] 
\label{Vlambdao}
\eea
The second term on the right can be interpreted as the Casimir energy contribution of all the spin zero states. The divergent sum over $n$ and integral over $k$ regulate each other provided we correlate the cut-offs via $k_{\rm max}^2 = n_{\rm max}^2\lambda$ and then remove the cut-off. 

Direct inspection of expression \eqref{ztot}-\eqref{zzero} for the total grand canonical partition function shows that it satisfies a $\Z_2$ strong-weak coupling duality symmetry that interchanges the discrete momenta and long string winding quantum numbers $m$ and $d$ and the
dimensionless $T\bar{T}$ coupling $\lambda$ with $1/\lambda$, provided that the vacuum energy contribution satisfies the following identity
\bea
V_0(\lambda) \is \lambda V_0(1/\lambda).
\eea
 We will now determine the constant $\mu_0$ by imposing this identity. First we use the Abel-Plana formula\footnote{The Abel-Plana formula states that for any function $f(t)$ that is holomorphic for $\Re(t)\geq 0$
$$
\sum_{n=0}^\infty f(n) - \int_0^\infty\!\!\! dk \, f(k) = \frac 1 2 f(0) + \i \int_0^\infty\!\!\!dt \; \frac{f(\i t)-f(-\i t)}{e^{2\pi t}-1}
$$
Setting $f(t) = \pi \lambda \sqrt{t^2 + 2E/\lambda}$ yields the identity
$$
 \sum_{n\in\mathbb{Z}} \pi \lambda  \sqrt{n^2 + {2E / \lambda}}   -  \int_{-\infty}^{ \infty}\!\! dk \, \pi \sqrt{k^2\!+\nspc 2E^{}}\, =  
 -  2 \! \int_{\strut{\sqrt{\nspc 2\ \;\;\;\,}\!\!\!\!\!\!\!\!\!  E\nspc/\nspc \lambda}}^{\infty}\!\! dp \, 2\pi \lambda \frac{\sqrt{p^2\! -\nspc 2E/\lambda}}{e^{2\pi p} -1} \quad
$$
 } to re-express $V_0(\lambda)$ as
\bea
V_0(\lambda) \is -\mu_0 
-  \nu_0(\lambda)
\eea
with $\nu_0(\lambda)$ defined in \eqref{nuzero}. Requiring the strong-weak duality symmetry
\bea
0= V_0(\lambda) - \lambda V_0(1/\lambda) \is -(1-\lambda) \mu_0 -  \bigl( \spc \nu_0(\lambda) - {\lambda} \spc \nu_0(1/\lambda)\spc \bigr)
\eea
reproduces the expression \eqref{munew} with $q$ set equal to $\lambda$.
The total vacuum energy contribution can thus be written in a manifestly $\Z_2$ duality-symmetric form as follows 
\bea
 {V_0(\lambda)} \is 
 \frac{\lambda}{1-\lambda}\bigl(\spc  \spc \nu_0(\lambda) - \spc \nu_0(1/\lambda)\spc \bigr)
\eea
We will discuss the convergence properties of this expression in the concluding section.

\subsection{Spectral properties of the $T\Bar{T}$-deformed free energy}
\label{sec:spectralpropsttbar}

Besides $O(2,2,\Z)$ T-duality symmetry and the triality between $\rho,\ssigma$ and $\ttau$, the $\Gamma_{2,2}$  Narain sum $\hatK(\rho,\ssigma,\ttau)$ satisfies a number of other remarkable properties. In particular, it satisfies the following relations
\bea
\label{deltaks}
	\Delta_{\ttau}\hatK(\rho,\ssigma,\ttau)\is \Delta_{\rho} \hatK(\rho,\ssigma,\ttau)=\Delta_{\ssigma} \hatK(\rho,\ssigma,\ttau), {}_{\strut}\label{laplaciank} \
\eea
where each laplacian acts on the corresponding argument. 
These relations can be verified directly from the explicit form of the Narain sum. This property was used in \cite{Benjamin:2021hcft} to derive an explicit formula, given in equation \eqref{SpectralGamma22}, for the spectral decomposition of $\hatK(\rho,\ssigma,\ttau)$ in terms of eigenfunctions of the three laplacians on the fundamental domain~${\cal F}$. Here we will use the result of  \cite{Benjamin:2021hcft}  to give an alternative definition of the $T\bar{T}$-deformed free energy $F(\rho,\ssigma)$ based on the spectral composition of the seed CFT partition function $Z_{c_0}(\ttau)$. 
The spectral function $E_s$ and $\nu_n$ are also eigenfunctions of the Hecke operators. The integration kernel $\hatK$ thus also satisfies the relation
\bea
\label{tjk}
\quad T_j^\ttau \hatK(\rho,\ssigma,\ttau)\is T_j^\rho \hatK(\rho,\ssigma,\ttau)=T_j^\ssigma \hatK(\rho,\ssigma,\ttau) \label{eq:heckettbaridentities1} \qquad \forall j
\eea
where each Hecke operator acts on the corresponding modular parameter via \eqref{heckeactt}. 

To obtain the spectral decomposition of $Z_{c_0}(\ttau)$, we need to find a way to regularize its divergent overlap integrals with the eigenfunctions of $\Delta_\ttau$ on the fundamental domain, the Eisenstein series $E_{s}(\ttau)$ and cusp forms $\nu_n(\tau)$.\footnote{Appendix \ref{sec:app22} list some properties of the Eisenstein series and cusp forms. For more details see \cite{cuspref}.} We will do this via the procedure outlined above: we will first compute these overlaps for the regulated seed partition function $Z_c(\ssigma)$ with negative central charge $c<0$ introduced in equation \eqref{newzc}. Since $Z_c(\ssigma)$ is regular near the cusp, its overlaps with the spectral functions give a finite result 
\bea
	 \int_{\mathcal{F}}\frac{d^2\ttau}{\ttau_2^2} Z_c(\ttau)E_{s}(\ttau)\is \beta_c(1-s) = \frac{\Lambda(1\nspc -\nspc s)}{\Lambda(s)}\beta_c(s),\\[2mm]
		\int_\mathcal{F} \frac{d^2\ttau}{\ttau_2^2} Z_c(\ttau) \nu_n(\ttau)\is \alpha_{c,n} (\nu_n,\nu_n),
\eea
with $\Lambda(s)$ defined in \eqref{lambdadef}.\footnote{
Here we assumed that $Z_c(\ttau)$ is an even function of $\ttau_1$, so it only has overlap with the even cusp forms. We suppress the + superscript in our notation.}  By construction, both overlaps are finite for $c<0$. Hence we deduce that in this regime $Z_c(\ttau)$ admits the Roelcke-Selberg spectral decomposition
\bea
\label{spectral}
		Z_c(\ttau) \is \varepsilon_c  +
		\frac{1}{4\pi \i }\int\limits_{\text{Re}s = \frac{1}{2}}\!\!\!\! ds \, \beta_c(s)\spc E_s(\ttau) + 
		\sum_{n=1}^\infty  \alpha_{c,n}\nu_n(\ttau).
\eea
Here $\varepsilon_c$ is related to the vacuum energy $\mu_c$ by via $\varepsilon_c  = 2\mu_c/{\rm vol({\cal F})}= {6\mu_c}/{\pi}$. At face value, this expansion and the coefficients $\varepsilon_c$, $\alpha_{c,n}$ and $\beta_c(s)$ are only well defined and finite in the regime $c<0$. However, it seems reasonable to assume that these coefficients are analytic functions of $c$ and can be uniquely analytically continued to the physical value $c=c_0$, even though $Z_{c_0}(\ssigma)$ itself is not square-integrable.

Combining the integral definition of the $T\bar{T}$-deformed free energy at general $c$ 
\bea
\label{fcint}
		F_c(\rho,\ssigma) \is \frac{1}{2}\int_{\mathcal{F}}\frac{d^2\ttau}{\ttau_2^2} \hatK(\ttau,\rho,\ssigma)Z_c(\ttau),
\eea	
with the spectral decomposition formula \eqref{SpectralGamma22} of the $\Gamma_{2,2}$ Narain partition function $\hatK(\rho,\ssigma,\ttau)$, we derive that the $T\bar{T}$-deformed free energy associated with the seed partition function \eqref{spectral} has the following spectral decomposition that pairs eigenfunctions of $\Delta_\tau$ with those of $\Delta_\rho$
\bea
\label{fdecomp}
		F_c(\rho,\ssigma) \hspace{-6mm} & & = \ \frac{\pi}{6} \varepsilon_c \, (\delta + \hat{E}_1(\rho)+\hat{E}_1(\ssigma)) \nn \\[-2mm]\\[-2mm]\nn
		& & +\ \frac{1}{4\pi \i}\int_{\cal C} ds\,\beta_c(s) \, \Lambda(s) E_s(\rho)E_s(\ssigma)  +\ 4\sum_{n}  \alpha_{c,n} \,\nu_n(\rho)\nu_n(\ssigma).
\eea
with $\delta = \frac{3}{\pi}(\gamma_E + \log 4\pi + 24\zeta'(-1) -2)$. The $\Z_2$ symmetry between $\ssigma$ and $\rho$ and modular {\rm PSL}(2,$\Z$) invariance our now both manifest.

We thus obtain a strikingly simple prescription for associating a $T\bar{T}$ deformed free energy to a seed CFT partition function: starting from the spectral decomposition of $Z_c(\ttau)$, we simply need to replace every mode function of $\ttau$ by a product of two identical mode functions of $\rho$  and $\ssigma$. From the spectral decomposition \eqref{fdecomp} we immediately see that $F_c(\rho,\ssigma)$ satisfies the property
\bea
\label{deltaf}
\Delta_\rho F_c(\rho,\ssigma)\is \Delta_\ssigma F_c(\rho,\ssigma).
\eea
This equation directly follows from the integral definition \eqref{fcint} and the property \eqref{deltaks} of the kernel ${\cal K}$. The analytical continuation of the central charge $c$ to the physical value $c_0$ will not violate this property. By virtue of the property \eqref{tjk} of the Narain kernel ${\cal K}$,
the $T\bar{T}$-deformed free energy furthermore satisfies
\bea
\quad T_j^\rho F_c(\rho,\ssigma)\is T_j^\ssigma F_c(\rho,\ssigma) \qquad \forall j_{\strut}.
\eea

\bigskip

\def\hatZ{Z}

\def\N{{\mathbb{N}}}
\section{Conclusion and summary of results}
\label{sec:conclusions}

\vspace{-1mm}

We have introduced an S-duality invariant extension of the $T\bar{T}$-deformed symmetric product CFT partition function. We presented evidence that this deformation is well defined for arbitrary seed CFTs with central charge $c\leq 6$. The free energy of the extended theory is defined by replacing the original wrapping number 1 integration kernel \eqref{ksumzero} that defines the standard $T\bar{T}$-deformed partition function with the full $\Gamma_{2,2}$ Narain partition sum \eqref{narain0} that includes all wrapping numbers. The partition function obtained via this procedure can be interpreted as a single-trace $T\bar{T}$-deformed partition function of symmetric product CFT, extended with an extra subsector associated with the zero spin spectrum $\{E_i, i \in {\cal S}\}$ of the CFT.

For a given seed CFT with energy and momentum spectrum $\{(E_i, j_i)\}$, the explicit formula for the deformed partition function reads as follows
\bea
\label{partfinal}
\boxed{\ \; \hatZ(\rho,\ssigma) = e^{-\rho_2 V_0(\lambda)} \prod_{d,m\in \Z}\spc \prod_{\substack{i | j_i=md}} \, 
\frac{1^{{}_{\strut}}}{1-p_1^{\spc d}\spc q_1^{\spc m} \spc e\raisebox{5pt}{\scriptsize $ -2\pi\ssigma_2\sqrt{ \left(d/\lambda\right)^2\! +\nspc m^2\! +\nspc 2E_i/ \lambda}$}{}_{\strut}}\ \; }
\eea
with $p_1 = e^{2\pi \i \rho_1}$, $q_1=e^{2\pi \i \ssigma_1}$, and $\lambda = {\ssigma_2}/{\rho_2}$. The vacuum energy contribution is~given~by 
\bea
\label{vzerot} 
 {V_0(\lambda)} \is  - \mu_0 - \nu_0(\lambda) = 
 \frac{\lambda}{1-\lambda}\bigl(\spc  \spc \nu_0(\lambda) - \spc \nu_0(1/\lambda)\spc \bigr)\\[3mm]
 \label{nuzerot}
 \nu_0(\lambda)\hspace{-7mm} & & = \,  2 \sum_{i\in\mathcal{S}} 
\!\! \int\limits_{{\;\;\;\mbox{\scriptsize $
\sqrt{2^{}\ \;\;\;\;\;\,}\!\!\!\!\!\!\!\!\!\!\!\! E_i/\nspc \lambda$}}
}^{\ \;\; \infty}\!\!\! dp \; {2\pi}\lambda\,  \frac{\sqrt{p^2\nspc - 2E_i^{\!}/ \lambda}}{e^{2\pi p} -1}\, 
 \eea
 where $\mu_0$ is a theory dependent constant given in equation \eqref{munew}.
The $d>0$ and $d<0$ subfactors in \eqref{partfinal} take the expected form of a deformed symmetric product CFT partition function, while the remaining $d=0$ subfactor represents the partition function of the zero wrapping sector. The combined formula \eqref{partfinal} takes the form of the trace over the Hilbert space of the deformed symmetric product CFT 
\bea
Z(\rho,\ssigma)  \is \tr_{{\cal H}}\bigl(e^{2\pi \i \hat{N}\rho} e^{ 2\pi \i (\ssigma_1 \hat{J} + \i\ssigma_2 \hat{H}{(\lambda)})}\bigr) 
\eea
where $\hat{N}$ counts the order of the symmetric product and $\hat{J}$ and ${\hat H(\lambda)}$ denote the momentum operator and the deformed Hamiltonian.
The form of the Hilbert space and the deformed energy spectrum can be read off from equation \eqref{partfinal}. As described in the previous sections, this spectrum looks like that of a second quantized string theory with worldsheet theory given by the product of the seed CFT and a gaussian sigma model with the two-torus with metric \eqref{torusm} as its target space. 

This non-zero wrapping factor with $d\neq 0$ in \eqref{partfinal} is well defined over the regime $c\leq 6$ and $\lambda <1$ in which all deformed energies are real. The zero wrapping factor with $d=0$ is also finite over the same parameter range
\bea
Z_0(\rho,\sigma) \equiv  e^{-\rho_2 V_0(\lambda)}
\prod_{i \in {\cal S}}  \prod_{n\in \Z} \, 
\frac{ 1}{1- e^{2\pi \i \ssigma_1 n -2\pi\ssigma_2 \nspc \sqrt{ n^2 + 2E_i/ \lambda}}}
 \,   \is\; e^{\rm finite}
\eea 
A more robust proof that the total partition function is well-defined is obtained by considering the free energy $F(\rho,\ssigma) = \log Z(\rho,\ssigma)$. It splits up as
\bea
F(\rho,\ssigma)  \is   F_0(\rho,\ssigma) +
{F}_+(\rho,\ssigma) + F_-(\rho,\ssigma)
\eea
where the respective terms represent the contribution of zero, positive, and negative wrapping number $d$.
The free energies ${F}_\pm$ are finite and real in our regime $c\leq 6, \lambda< 1$. 
The first term $F_0(\rho,\ssigma) = \log Z_0(\rho,\ssigma)$ requires more careful consideration. We have shown that it can be written as an infinite sum of Bessel functions as follows
\begin{align}
\label{finalone}
F_0(\rho,\ssigma) &=  \mu_0 \rho_2  +  \sum_{i\in\mathcal{S}}\, \sum_{(m,n)\neq (0,0)} \sqrt{\frac{2 \rho_2 \ssigma_2 E_i}{|m\ssigma+n|^2}}\, K_1\left(\sqrt{\frac{8\pi^2 \rho_2 |m\ssigma+n|^2 E_i}{\ssigma_2}}\right)\\[5mm]
\mu_0 &=  
\sum_{i\in\mathcal{S}} 2\sqrt{2E_i} \sum_{m=1}^\infty \left[\frac1m K_1\bigl(2\pi m \sqrt{2 E_i}\bigr) + 2\pi\sqrt{2E_i} K_0(2\pi m \sqrt{2E_i}\bigr)\right]
\label{finaltwo}
\end{align}
We have numerically evaluated this expression for various seed CFTs and found that it is finite and satisfies a number of remarkable properties listed below.
 In Appendix \ref{sec:numerics}, we explain more details of the numerical tests.

By virtue of the $T$-duality symmetry and spectral characteristics of the Narain partition function, we find that the deformed free energy $F(\rho,\ssigma)$ and partition function $Z(\rho,\ssigma)$ exhibits all the following symmetries:

\bigskip

\fbox{
    \parbox{14cm}{
\begin{itemize}
\item{Modular symmetry:}~~~~~~~~~~~~~~~~~~~~~\,$F(\rho,\ssigma) \, = \; F\bigl(\rho,\mbox{\large $\frac{{a} \ssigma +{b} } {{c} \ssigma + {d}}$}\bigr)$, 
\medskip
\item {Mirror symmetry:}~~~~~~~~~~~~~~~~~~~~~~~~\spc$F(\rho,\ssigma) \, =\; F(\ssigma,\rho)$,
\medskip
\item  {S-duality symmetry:}~~~~~~~~~~~~~~~~~~~~~$F(\rho,\ssigma) \, = \; F\bigl(\mbox{\large $\frac{{a} \rho +{b} } {{c} \rho + {d}}$},\ssigma\bigr)$,
\medskip
\item  {Spectral symmetry:}~~~~~~~~~~~~~~~~~~~~$\Delta_\rho \, F(\rho,\ssigma) \, = \, \Delta_\ssigma\, F(\rho,\ssigma)$,
\medskip
\smallskip
\item {Hecke symmetry:}~~~~~~~~~~~~~~~~~~~~~~~\,${T}_j^\rho \, F(\rho,\ssigma) \, = \, {T}_j^\ssigma \, F(\rho,\ssigma)$,
\medskip
\smallskip
\item {U-duality symmetry:}~~~~~~~~~${\rm O}(2,2;\mathbb{Z})\simeq {\rm{\rm PSL}}(2,\mathbb{Z})\times {\rm{\rm PSL}}(2,\mathbb{Z})\rtimes \mathbb{Z}^2_2$
\end{itemize}}}

\bigskip

\noindent
From a mathematical point of view, it remains remarkable statement that one can associate to any CFT partition function with $c\leq 6$ a new deformed free energy and partition function  with all the above properties.

From a physical point of view, the extended ${\rm O}(2,2;\mathbb{Z})$ U-duality symmetry follows from the combination of (i) modular invariance under {\rm PSL}(2,$\mathbb Z$) transformations acting on the torus modulus $\ssigma$ with (ii) the $\Z_2$ mirror symmetry that interchanges complex coupling $\rho$ with $\ssigma$. The $\Z_2$ mirror symmetry acts on the Hilbert space ${\cal H}$ of the symmetric product CFT by interchanging the momentum quantum number $m$ with the long string winding number $d$. The mirror symmetry is a manifest property of the final expression \eqref{partfinal}-\eqref{nuzerot} for $Z(\rho,\ssigma)$ thanks in particular to the identity $
V_0(\lambda) = \lambda V_0(1/\lambda).$

 We end with some comments about the potential physical significance of our results.

\medskip

\noindent
{\it Non-perturbative completion}

\smallskip

To obtain the duality invariant partition function, we had to include the second factor in \eqref{partfinal} given by the partition function of the spinless particles of mass squared $2E_i/\lambda$, in units of the torus radius.  To what extent does adding this sector amount to a non-perturbative completion of the conventional $T\bar{T}$-deformation? First we note that in the $\lambda$ to zero limit, all the  spin zero particles become infinitely massive and therefore decouple. More concretely, the free energy $F_0(\rho,\ssigma)$ of the extra sector has a series expansion in inverse powers of $\lambda$. Therefore the manifestation of the extra sector cannot be seen in perturbation theory in $\lambda$. 

The contribution from the negative-wrapping sectors also has a non-perturbative nature\footnote{We thank Shota Komatsu for pointing this out to us.}. The deformed-energy of negative-wrapping sectors scales for small $\lambda$ as $|d|/\lambda$, so this sector does not contribute to the perturbative expansion in $\lambda$.  It would be interesting to see if one can derive the non-perturbative completion of this series by treating the perturbative part of the $T\bar{T}$-deformed partition function as an asymptotic series in $\lambda$ and applying resurgence.

\medskip

\noindent
{\it Eliminating complex energies}

\smallskip

It is natural to view the duality symmetry as a gauge symmetry, i.e. as a true physical identification between systems related via a duality transformation. The duality gauge symmetry has the benefit that it eliminates one of the seemingly pathological features of the $T\bar{T}$ deformed theory, namely the appearance of complex energies for strong $T\bar{T}$-couplings $\lambda>6/c$. The argument is simple: the mirror symmetry that interchanges $\rho$ and $\ssigma$ is a strong-weak duality that maps $\lambda = \ssigma_2/\rho_2$ to $1/\lambda = \rho_2/\ssigma_2$. Hence it is always possible to go to a duality frame in which $\lambda\leq 1$. This is sufficient to avoid the appearance of complex energies, as long as $c\leq 6$. 

This implication of the strong-weak duality property is an indication that our extension of the $T\bar{T}$-deformed CFT for $c\leq 6$ may define a UV complete theory. The second quantized theory given by the sum over all wrapping sectors appears to be better behaved than the first quantized theory given by an individual wrapping sector.
This looks somewhat coincidental. However, it is somewhat reminiscent of the resolution of the Klein paradox via the second quantized interpretation of the Dirac equation. Indeed, the appearance of complex energies in $T\bar{T}$-deformed CFT may be telling us that the deformed theory should not be formulated as the worldsheet of a single string but rather as the collective worldsheet description of a second quantized string theory. Our results thus provide support to the proposed interpretation of $T\bar{T}$-deformed CFT in terms of the holographic dual of little string theory\cite{Giveon:2017nie,Giveon:2017myj,Apolo:2019zai,Chakraborty:2019mdf}.

\medskip
\medskip

\noindent
{\it Spectral definition of the $T\bar{T}$ deformation}

\smallskip

In section \ref{sec:spectralpropsttbar}, we found that the $T\bar{T}$-deformation can be formulated as a direct map from the spectral decomposition \eqref{spectral} of the seed CFT partition function $Z_{\rm CFT}(\tau)$ to the spectral decomposition \eqref{fdecomp} of the deformed free energy of the symmetric product CFT. In essence, the deformation amounts to the replacement of the eigenfunctions of the laplacian on the fundamental $\ttau$ domain by the product of the same eigenfunctions on the fundamental domains of $\rho$ and $\ssigma$.  This spectral characterization of the $T\bar{T}$-deformed free energy  hints at a deeper geometric significance of our results. 

The spectral decomposition of the $T\bar T$-deformation allows us to also formally average the (complexified) $T\bar T$ coupling over the fundamental domain $\mathbb{H}/SL(2,\mathbb Z)$. Although this average diverges for the same reason the moduli space average of the $c=2$ Narain partition function diverges (see \cite{Afkhami-Jeddi:2020ezh, Maloney:2020nni}), we can formally define an average by reading off the constant piece in the spectral decomposition.

    More explicitly,
    \begin{align}
        \langle F(\rho, \tau) \rangle_\rho &= \frac{\frac12 \int_{\mathcal{F}} \frac{d^2\rho}{\rho_2^2} \int_{\mathcal{F}} \frac{d^2\sigma}{\sigma_2^2} \mathcal{K}(\rho, \tau, \sigma) Z(\sigma)}{\frac \pi 3}. 
    \end{align}
    The integral $\int_{\mathcal{F}} \frac{d^2\rho}{\rho_2^2} \mathcal{K}(\rho, \tau,  \sigma)$ diverges, but we define it via its spectral decomposition in (\ref{SpectralGamma22}) to be $\hat E_1(\sigma) + \hat E_1(\tau) + \frac3\pi\left(-2 + \gamma_E + 24 \zeta'(-1) + \log(4\pi)\right)$. We then get
      \begin{align}
        \langle F(\rho, \tau) \rangle_\rho &= \frac{3}{2\pi} \int_{\mathcal{F}} \frac{d^2\sigma}{\sigma_2^2} \left[\hat E_1(\sigma) + \hat E_1(\tau) + \frac3\pi\left(-2 + \gamma_E + 24 \zeta'(-1) + \log(4\pi)\right)\right] Z(\sigma) \nonumber \\
        &= \frac3\pi \mu_0 \left[\hat E_1(\tau) + \frac3\pi\left(-2 + \gamma_E + 24 \zeta'(-1) + \log(4\pi)\right)\right] + \frac{3}{2\pi} \int_{\mathcal{F}} \frac{d^2\sigma}{\sigma_2^2} \hat E_1(\sigma) Z(\sigma) \nonumber \\
        &= -\frac{18}{\pi^2} \mu_0 \left[\log(\sqrt{y} |\eta(\tau)|^2)\right] + \frac{9}{\pi^2}\mu_0(\gamma_E - \log(4\pi)) + \frac{3}{2\pi} \int_{\mathcal{F}} \frac{d^2\sigma}{\sigma_2^2} \hat E_1(\sigma) Z(\sigma).
    \end{align}
    Thus we have
    \begin{align}
        e^{\langle F(\rho,\tau)\rangle_{\rho}} \propto y^{-\frac{9
    \mu_0}{\pi^2}} |\eta(\tau)|^{-\frac{18\mu_0}{\pi^2}}.
    \end{align} 

\medskip
\medskip

\noindent
{\it Wheeler-DeWitt equation}

\smallskip

It can be shown that the spectral symmetry equation  $(\Delta_\rho -\Delta_\ssigma) F(\rho,\ssigma)=0$ for the free energy takes the same form as the Wheeler-DeWitt equation satisfied by the mini-superspace wavefunction in $AdS_3$ gravity, extended by including an anti-symmetric B-field, defined on a three manifold given by a spatial two torus $\mathbb{T}^2$ times time. This correspondence is compatible with the proposed holographic interpretation of the $T\bar{T}$ deformed theory with $\lambda<0$ as the boundary dual of AdS gravity with a finite radial cut-off. Note, however, that taking $\lambda<0$ reintroduces the property that the deformed energies become complex and takes us outside of the regime of validity of our analysis. 

\medskip
\medskip

\noindent{\it More general deformations}

\smallskip
Finally, we point out that our S-duality invariant extension of the $T\bar{T}$ deformation may be regarded as a specific instantiation of a more general procedure that generates a family of modular invariants from a seed modular-invariant partition function.\footnote{This can be thought of as a generalization of the $f(H)$ deformations considered in \cite{Gross:2019uxi}.} 

Consider the following integral of the partition function over the fundamental domain of ${\rm PSL}(2,\mathbb{Z})$
\begin{equation}\label{eq:generalDeformedZ}
    Z[\varphi](\tau) = \int_{\mathcal{F}} {d^2\sigma\over \sigma_2^2}\mathcal{K}[\varphi](\tau,\sigma)Z_{\rm CFT}(\sigma)
\end{equation}
where the kernel $\mathcal{K}[\varphi]$ is given by a sum over ${\rm PSL}(2,\mathbb{Z})$ images
\begin{equation}
    \mathcal{K}[\varphi](\tau,\sigma) = \sum_{\gamma\in {\rm PSL}(2,\mathbb{Z})} \mathcal{K}_0[\varphi](\tau,\gamma\sigma).
\end{equation}
Here the seed $\mathcal{K}_0[\varphi]$ is a function only of the geodesic distance between the points $\tau$ and $\sigma$ on the upper half-plane
\begin{equation}
    \mathcal{K}_0[\varphi](\tau,\sigma) = \varphi\left({|\tau -\bar\sigma|^2\over \tau_2\sigma_2}\right).
\end{equation}
The off-shell gravitational path integral of AdS$_3$ Einstein gravity on the Euclidean wormhole with the topology of a torus times interval \cite{Cotler:2020ugk,Cotler:2020hgz} is an example of such a deformation kernel, with $\varphi(x) = 1/x$. Since the seed kernel $\mathcal{K}_0[\varphi]$ is invariant under modular transformations that act simultaneously on $\tau$ and $\sigma$, the deformed partition function defined by (\ref{eq:generalDeformedZ}) is a modular-invariant function of $\tau$, provided that the integral is convergent or appropriately regularized.

\bigskip

\section*{Acknowledgments}

We thank Shazia-Ayn Babul, Cyuan-Han Chang, Liam Fitzpatrick, Shota Komatsu, Alexander Maloney, and Erik Verlinde for very valuable discussions. The research of NB is supported by the Sherman Fairchild Foundation and the U.S. Department of Energy, Office of Science, Office of High Energy Physics Award Number DE-SC0011632. The research of SC is supported by the Sam B. Treiman fellowship at the Princeton Center for Theoretical Science. The research of HV is supported by NSF grant PHY-2209997. JK is supported by NSF grant PHY-2207584.

	\appendix
	
\section{Functional determinant via the heat kernel method}

\vspace{-2mm}

\label{sec:heatkernel}
In the main text we make use of the heat kernel on the two torus and its relation to the functional determinant. Suppose we want to compute the free energy 
\bea
F(\tau, M) \is - \frac{1}{2}\log\bigl(\det(-\Delta + M^2)\bigr),
\eea
of a free single-valued real scalar field $\phi$ with action 
\bea
	S[\phi] = \frac{1}{2}\int_{\mathbb{T}^2}\!\! d^2 x \sqrt{g}\, \phi (-\Delta + M^2) \phi.
\eea 
defined on the two torus with metric \eqref{torusm}. Assuming that $M^2>0$, the partition function is given in terms of the determinant of the kinetic operator $-\Delta + M^2$. This determinant can be calculated as follows. 
	
The heat kernel $K(t,x,y)$ is defined as the unique solution to the heat equation 
\bea
	\bigl(\frac{\partial}{\partial t}+\Delta_x\bigr)K_0(t,x,y) = 0
\eea
with the  initial condition $K_0(0,x,y) =\delta^{(2)}(x-y)$. The one-loop determinant on $\mathbb{T}^2$ can be written in terms of the heat kernel as 
\bea
	-\log \det\bigl(-\Delta + M^2\bigr) \is \int_0^\infty \frac{dt}{t}\, e^{-M^2t}\int\! d^2x \sqrt{g} \, K_0(t,x,x)\,  \nn\\[-2mm]\\[-2mm]\nn
	\is (2\pi)^2\rho_2 \int_0^\infty \frac{dt}{t} \, K_0(t,x,x)\, e^{-M^2 t}.
\eea

On $\mathbb{R}^2 \simeq \mathbb{C}$,  the heat kernel takes the familiar form $K_0^{\mathbb{R}^2}(t,x,y) = \frac{1}{4\pi t}\, \exp\bigl(-\frac{|x-y|^2}{4t}\bigr).$
We can compactify $\mathbb{R}^2$ to $\mathbb{T}^2$ by enforcing periodic boundary conditions via the method of images: we identify $\mathbb{T}^2 = \mathbb{C}/(\mathbb{Z}+\mathbb{Z}\, \ssigma)$ and sum over all $\mathbb{Z}\oplus \mathbb{Z}\, \ssigma$ images. Using two integers $(c,d)$ to label the images, the heat kernel on the torus with metric \eqref{torusm} can be written in terms of a sum
\bea
	\label{heattrace}
	K_{0}(t,x,x) \is \sum_{c,d\in\mathbb{Z}}\frac{1}{4\pi t}\exp\Bigl({-\frac{4\pi^2 \rho_2 |d+c\ssigma|^2}{4t\ssigma_2}}\Bigr).
\eea
Note that to any pair $(c,d)$ except $(0,0)$ we can associate an element $\gamma$ in the group $\Gamma_\infty \backslash{\rm PSL}(2,\mathbb{Z})$. Setting $\text{gcd}(c,d) = r$ and $c>0$, we can always find an element $\gamma = \bigl({\footnotesize \!\begin{array}{cc}\!
			a \! &\! b\\[-1mm]
			{c}/{r}\! &\! {d}/{r}\!\end{array}}\!\bigr)$ with $\det \gamma = 1$. The group $\Gamma_\infty \backslash{\rm PSL}(2,\mathbb{Z})$ transforms the complex structure of the torus $\ssigma \rightarrow \gamma \ttau$. Identifying ${|d+c\ssigma|^2}/{\ssigma_2}$ in \eqref{heattrace} with ${r^2}/{(\gamma \ssigma)_2}$ we can write the heat kernel on the torus as a Poincar\'{e} series
\bea
	\label{HeatKernel}
	K_{0}(t,x,x) =\frac{1}{4\pi t}+2\sum_{r= 1}^\infty\sum_{\gamma \in \Gamma_{\infty}\!\nspc\backslash{\rm PSL}(2,\mathbb{Z})} \frac{1}{4\pi t}\, \exp\Bigl({-\frac{4\pi^2 \rho_2  r^2}{4t\spc (\gamma \ssigma)_2}}\Bigr).
\eea
The factor of $2$ in front of the sum accounts for the $d<0$ terms in the $(c,d)$ summation. 
\def\twoE{M^2}	

\section{Functional determinant as a thermal partition function}
\label{sec:funcdet}
\def\tM{\,\widetilde{\!\!M\nspc}{\spc}}

In this Appendix we show how the free energy $F(\tau, M) = - \frac 1 2 \log \det(-\Delta+M^2)$ of a massive scalar field on the two torus can be recast in the form of a quantum mechanical partition function of the form of a trace over a Hilbert space. We will follow a modified version of the calculation described in \cite{iz} for the case of a massless scalar field.

The set of eigenvalues of $-\Delta + M^2$ on a two torus with metric \eqref{torusm} are given by  
\bea
\lambda_{n,m} = \frac{|m+n\tau|^2}{\tau_2\rho_2} + M^2
\eea 
with $n,m \in \Z$. We can obtain the functional determinant $\det(-\Delta + M^2) =  \prod_{n,m} \lambda_{n,m}$ through analytic continuation via the meromorphic function $G(s)$ defined as follows
\bea
\label{fgprime}
F(\tau, M) \is  \frac{1}{2}G'(0),\qquad \qquad
	G(s) =\sum_{m,n \in \Z} \;\frac{1}{\bigl( \frac{|m+n\tau|^2}{\tau_2\rho_2}+ \twoE\bigr)^{\! s}}
\eea
Since, due to the sum over all integers $m$, $G(s)$ is periodic function of $n\tau_1$ with unit period, we can Fourier expand and compute
\bea
	\sum_m \frac{1}{(\frac{|m+n\tau_2|^2}{\rho_2\tau_2}  + \twoE){\strut}^s}  \is \sum_\ell e^{2\pi \i \ell n \tau_1} \int_0^1 \!\! dy \, e^{-2\pi \i \ell y} \sum_m \frac{1}{(\frac{|m+y+\i n\tau_2|^2}{\rho_2\tau_2} + \twoE){\strut}^s}\nn \\[2mm]
	\is \sum_\ell \int_{-\infty}^\infty\!\! dy \, e^{2\pi \i \ell(n\tau_1 - y)} \frac{1}{(\frac{|y+\i n\tau_2|^2}{\rho_2\tau_2} + \twoE){\strut}^s} \nn\\[-2mm]\label{beefour}\\[-2mm]
	\is \frac{1}{\Gamma(s)} \sum_\ell \int_{-\infty}^\infty\!\! dy \, \int_0^\infty \!\!dt\;  t^{s-1} e^{2\pi \i \ell(n\tau_1 - y) -t\mbox{\footnotesize $ (\frac{y^2 + n^2 \tau_2^2}{\rho_2\tau_2}$} + \twoE)}\nn\\[2mm]\nn
	\is \frac{(\pi \rho_2\tau_2)^{1/2}\!\!}{\Gamma(s)}\,\;  \sum_\ell \int_0^\infty \!\!dt \; t^{s-3/2} e^{2\pi \i \ell n\tau_1 - t(\mbox{\footnotesize $\frac{n^2 \tau_2}{\rho_2}$} + \twoE) - {\pi^2 \rho_2 \tau_2 \ell^2}/{t}}
\eea
The integral for $\ell=0$ gives $(n^2 \frac{\tau_2}{\rho_2}  + M^2)^{\frac 1 2 -s} \Gamma(s-\frac 1 2 )$. The remaining terms can be evaluated via 
\bea
\int_0^\infty \! \frac{dt}{t^{3/2}} \, e^{-At  - B/t} \is \sqrt{\frac{\pi}{B}} e^{-2\sqrt{AB}}
\eea
We thus find that as $s$ approaches 0
\bea
     G(s)\is -1  -2\pi s \sum_n \,\tau_2{\textstyle\sqrt{n^2 \! + \frac{\rho_2 \twoE}{\tau_2}}}+  s \sum_\ell{\strut}' \sum_n \frac{1}{|\ell|} \, e^{2\pi \i \ell n\tau_1 -2\pi |\ell| \tau_2 \sqrt{n^2 \! + \frac{\rho_2 \twoE}{\tau_2}}}
\eea
This gives as our final result for the free energy \eqref{fgprime}
\bea
{F(\tau,M)} \is  -\!  \sum_n \pi \tau_2 {\textstyle\sqrt{n^2\! +\frac{\rho_2 \twoE}{\tau_2}}}
 - \sum_n  \log\Bigl(\nspc1\!-\nspc e^{2\pi \i n\tau_1 -2\pi \tau_2\sqrt{n^2 \! + \frac{\rho_2 \twoE}{\tau_2}}}\Bigr)\ \ 
\eea
The associated partition function ${Z(\tau,M)} = e^{F(\tau,M)}$ takes the form of a trace over the Hilbert space of a free  scalar field.

Performing a Poisson resummation on the sum over $n$ in the last line in \eqref{beefour} and taking the derivative at $s=0$ gives
\bea
F(\tau,M) \is -{\rho_2}  \int_{-\infty}^\infty \!\! dk \, \pi \sqrt{k^2 + M^2} + \pi \rho_2 \sum_{(n,\ell) \neq (0,0) } \int  \frac{dt}{t^2} \, e^{-t M^2  - \frac{\pi^2 \rho_2}{t\tau_2} |n+ \ell\tau|^2} 
\eea
Via the identification $M^2 = 2E_i$, the discrete sum over $n$ and $\ell$ can be recognized as a term in the Bessel function expression \eqref{fzerotwo} for the free energy $F_0(\rho,\sigma)$.

\section{Spectral decomposition of the $\Gamma_{2,2}$ Narain sum}
\label{sec:app22}
\vspace{-2mm}

The Eisenstein series $E_s$ is the real-analytic modular form defined by the meromorphic continuation of the following sum
\bea
	E_s(\tau) \is \sum_{\gamma \in \Gamma_{\infty}\backslash{\rm PSL}(2,\mathbb{Z})} \text{Im}(\gamma\tau)^s.
\eea
It is the simultaneous eigenfunction of both the Laplacian $\Delta_{\ttau}$ on the upper-half-plane 
and of the Hecke operators  $T_j$ with the eigenvalues
\bea
	\Delta_\tau E_s(\tau) \is s(1-s)E_s(\tau), \\[2mm]
	T_j E_s(\tau) \is 	\frac{\sigma_{2s-1}(j)}{j^{s-\frac{1}{2}}}E_s(\tau),
\eea
where $\sigma_n(j) = \sum_{d|j}d^n$. Here $T_j$ is defined as
\bea
	\label{heckeactt}
	{T}_j f(\tau) \is \frac{1}{\sqrt{j}} \sum_{\substack{ad = j,\, d>0 \\ b \,{\rm mod}\, d}} f\Bigl(\frac{a\tau+b}{d}\Bigr).
\eea
The Eisenstein series constitute the continuous eigenspectrum of $\Delta_\tau$ and $T_j$.
	
The cusp form $\nu_n^{\pm}(\tau)$ are the eigenfunctions of $\Delta_{\tau}$ associated with the discrete part of the eigenspectrum.
They are also eigenfunctions of the Hecke operators $T_j$
\bea
		\Delta_\tau \nu^\epsilon_n(\tau)\is \left(\frac{1}{4}+(R_n^\epsilon)^2\right)\nu_n^\epsilon(\tau)\\[2mm]
		 T_j^\tau \nu_n^\epsilon(\tau) \is a_j^{n,\epsilon}\nu_n^\epsilon(\tau),
\eea
where $a_j^{n,\epsilon}$ and $R_n^\epsilon$ are associated with the Fourier decomposition of the cusp form via
	\begin{align}
		\nu_n^+ (\tau) &= \sum_{j=1}^\infty a_j^{n,+} \cos{(2\pi j x)}\sqrt{y}K_{\i R_n^+}(2\pi jy) \nonumber \\ 
		\nu_n^- (\tau) &= \sum_{j=1}^\infty a_j^{n,-} \sin{(2\pi j x)}\sqrt{y}K_{\i R_n^-}(2\pi jy),
		\label{eq:cuspdefnsexp}
	\end{align}
where $\tau = x+\i y$ and $K_a(x)$ is the Bessel-K function.  The superscript $\epsilon = \pm$ labels the parity of the cusp forms
\bea
		\nu^+(\tau) = \nu^+(-\bar{\tau}),\quad \nu^-(\tau) = -\nu^-(-\bar{\tau}),
\eea
which can be readily seen from the Fourier decompositions (\ref{eq:cuspdefnsexp}).

Following the analysis in \cite{Benjamin:2021hcft}, we can perform a spectral decomposition of the $c=2$ Narain primary partition function as follows
\bea
	\label{SpectralGamma22}
	\hatK(\rho,\ssigma,\ttau) \is \alpha + \hat{E}_1(\rho) \nspc + \nspc \hat{E}_1(\ssigma)\nspc +\nspc \hat{E}_1(\ttau) \nspc + \nspc 
	\frac{1}{4\pi \i}\! \int_{{\cal C}}\! ds \spc \frac{2 \Lambda(s)^2}{\Lambda(1\!-\!s)}E_s(\rho)E_s(\ssigma)E_s(\ttau)\nonumber\\
	&&\qquad\quad \ +\; 8\spc 
	\sum_{\epsilon=\pm}\sum_{n=1}^\infty\, \delta_\epsilon\, \frac{\nu_n^\epsilon(\rho)\spc \nu_{n}^\epsilon(\ssigma)\spc \nu_{n}^\epsilon(\ttau)} {(\nu_n^\epsilon,\nu_n^\epsilon)}.
\eea 
The contour $\mathcal C$ is given by $\Re \spc s = \frac 12$. The remaining functions are defined as $\Lambda(s) \equiv \pi^{-s}\Gamma(s)\zeta(2s)$, $\hat{E}_1$ represents the non-singular part of $E_1$,  and $\alpha$ is a moduli-independent constant:
\bea
	\label{hatE}
	\hat{E}_1 \, \equiv \,  \lim_{s\rightarrow1}\Bigl(E_s\! - \! \frac{3}{\pi (s-1)}\Bigr),\qquad \quad
	\alpha\equiv \frac{3}{\pi}(\gamma_E+3\log(4\pi)+48 \zeta'(-1)-4).
\eea
The $\epsilon = \pm$ superscript labels two parities of the cusp forms and we have $\delta_{+} = 1,\delta_{-} = -\i$.  In the spectral decomposition expression, the triality symmetry among three moduli parameters $\rho,\ssigma,\ttau$ is manifest. 
	
By virtue of its spectral decomposition (\ref{SpectralGamma22}), we deduce the $\hatK$ satisfies the identities
\bea
		\Delta_{\ttau}\hatK(\rho,\ssigma,\ttau)\is \Delta_{\rho} \hatK(\rho,\ssigma,\ttau)=\Delta_{\ssigma} \hatK(\rho,\ssigma,\ttau), \label{eq:laplacianttbaridentities} \\[4mm]
		T_j^\ttau \hatK(\rho,\ssigma,\ttau)\is T_j^\rho \hatK(\rho,\ssigma,\ttau)=T_j^\ssigma \hatK(\rho,\ssigma,\ttau).
		\label{eq:heckettbaridentities}
\eea
where each Hecke operator acts on the corresponding modular parameter. The relations \eqref{eq:laplacianttbaridentities} can also be verified directly from the explicit form and the triality symmetry of the Narain sum. The function $\hat E_1$ defined in equation \eqref{hatE}  is not an eigenfunction of the Laplacian and Hecke operators. Rather, it has an inhomogeneous term
 \bea
        \Delta_\ttau \hat E_1(\ttau) \is -\frac3\pi \qquad \quad
        T_j^\ttau \hat E_1(\ttau) = \frac{\sigma_1(j)}{\sqrt j} \hat E_1(\ttau) + \#,
        \label{eq:heckelaplacianstuff}
\eea
where $\#$ is a $\ttau$-independent constant. Combining (\ref{eq:heckelaplacianstuff}) with the fact that $T_j^\ttau 1 = \frac{\sigma_1(j)}{\sqrt j}$ we obtain (\ref{eq:heckettbaridentities}).
    
\section{Regularized integral of $Z_{\rm CFT}(\ttau)$}
	\label{sec:integralregapp}
\vspace{-1mm}
 In this Appendix, we look at the integral of a partition function $Z_{\rm CFT}(\ttau)$ over the fundamental domain $\mathcal{F}$. Motivated by Narain's family of free boson CFTs, we first look at the following modified partition function:
\begin{equation}
		\hat Z_{\rm CFT}(\ttau,d) \equiv y^{d/2} |\eta(\ttau)|^{2d} Z_{\rm CFT}(\ttau).
		\label{eq:z1taudef}
\end{equation}
For Narain CFT, when $d =c$ (the central charge), we have $\hat{Z}_{\rm CFT}(\ttau,c)$ as the $U(1)^c$ primary-counting partition function. In order to compute the integral of $\hat{Z}_{\rm CFT}(\ttau,c)$ over $\mathcal{F}$, we consider the crossing equation for scalar operators for a Narain CFT at central charge $c$, found in (3.22) of \cite{Benjamin:2022pnx}:
\begin{align}
		\sum_{i \in \mathcal{S}} e^{-2\pi E_i y} &= \frac{\Lambda\left(\frac{c-1}2\right)}{\Lambda\left(\frac c2\right)} y^{1-c} + \varepsilon_c y^{-\frac c2} + \sum_{k=1}^\infty \text{Re}\left(\delta_{k,c} y^{-\frac c2+1-\frac{z_k}2}\right) \nonumber \\&+ \frac{y^{1-c}}{\sqrt \pi} \sum_{\substack{i \in \mathcal S\\ i \neq \text{vac}}}\sum_{n=1}^\infty b(n) n^{c-2} U\left(-\frac12, \frac c2, \frac{2\pi n^2 E_i}y\right) e^{-\frac{2\pi n^2 E_i}y},
		\label{eq:crossing}
\end{align}
where $\mathcal{S}$ is the set of scaling dimensions for all scalar $U(1)^c$ primary operators of our CFT, and the second line of (\ref{eq:crossing}) does not include the vacuum operator. Note also that because we multiplied by $|\eta(\sigma)|^{2c}$ in (\ref{eq:z1taudef}), the $E_i$'s in (\ref{eq:crossing}) do not include the Casimir shift of $-\frac c{12}$. The other terms are defined as:
\bea
	\label{lambdadef}
		b(n) :\!\!\is \sum_{k|n} k \mu(k)  \qquad \qquad \Lambda(s) := \pi^{-s} \Gamma(s)\zeta(2s),
\eea
$U$ is the confluent hypergeometric function of the second kind, $\mu$ is the M\"obius function, the constants $z_k$ are the nontrivial zeros of the Riemann zeta function (with positive imaginary part), and $\delta_{k,c}$ are theory-dependent constants. Finally and most importantly $\varepsilon_c$ is the (theory-dependent) constant piece in the spectral decomposition:
\begin{equation}
		\varepsilon_c = \lim_{d\rightarrow c} \int_{\mathcal{F}} \frac{dx dy}{y^2} \frac{\hat Z_{\rm CFT}(\ttau,d) }{\pi/3},
		\label{eq:varepsdef}
\end{equation}
with $\hat Z_{\rm CFT}(\ttau,d)$ defined as in (\ref{eq:z1taudef}).
We find that $\varepsilon_c$ is essentially the integral we want to compute and it comes back to our original partition function integral when $d = 0$ as opposed to $d=c$. In particular comparing with (\ref{vacenergy})
\begin{equation}
    \varepsilon_0 = \frac{6\mu_0}{\pi}.
\end{equation}

Our main strategy in this section will be to apply a cleverly chosen linear functional to isolate the term $\varepsilon_c$ as a convergent sum in terms of the scalar operator scaling dimensions of the theory (i.e. the set $\mathcal{S}$).
	
Let us relabel $y \equiv t^{-2}$ and rewrite (\ref{eq:crossing}) as
\begin{align}
		&- \sum_{i \in \mathcal{S}} t^{2-c} e^{-\frac{2\pi E_i}{t^2}} + \frac{\Lambda\left(\frac{c-1}2\right)}{\Lambda\left(\frac c2\right)} t^{c} + \varepsilon_c t^2 + \sum_{k=1}^\infty \text{Re}\left(\delta_{k,c} t^{z_k}\right) \nonumber \\& ~~~~~~~~~~ + \frac{t^{2-c}}{\sqrt \pi} \sum_{\substack{i \in \mathcal S \\ i \neq \text{vac}}}\sum_{n=1}^\infty b(n) n^{c-2} U\left(-\frac12, \frac c2, 2\pi n^2 E_i t^2\right) e^{-2\pi n^2 E_i t^2} = 0.
		\label{eq:asdf}
\end{align}
Now let us apply a linear functional to this. Define
\bea
	\Phi_q(t) \equiv \sum_{n=1}^\infty \varphi_q(nt) = \sum_{n=1}^\infty e^{-\pi q n^2 t^2}.
	\label{eq:varphidef}
\eea
The functional we apply is defined as follows\footnote{Strictly speaking we need to take a linear combination of the functionals defined here such that both $\varphi(t)$ and its Fourier transform vanish at $t=0$ in order for the functional to make sense. However, this implies that a single Gaussian for $\varphi(t)$ as in (\ref{eq:varphidef}) can only allow terms proportional to a constant and $q^{-1/2}$. See Sec 3.3 of \cite{Benjamin:2022pnx} for more detail.\label{footnotedanylo}}.
\begin{align}
		\mathcal{F}_q[h(t)] \equiv \int_0^\infty \frac{dt}{t} h(t) \Phi_q(t).    
		\label{eq:functionaldanylodef}
\end{align}
The reason we choose this functional is that it is designed to kill the sign-indefinite $\delta_{k,c}$ terms in the crossing equation. In particular, the first three terms are simple, using (3.32) of \cite{Benjamin:2022pnx}:
\bea
	\mathcal{F}\left[t^s\right] = \frac12 q^{-\frac s2} \Lambda\left(\frac s2\right).
\eea
By design, this causes the terms proportional to $t^{z_k}$ to vanish. Finally, following the discussion in footnote \ref{footnotedanylo} (and the end of Sec 3.3 of \cite{Benjamin:2022pnx}), we know the RHS after applying this functional must be $c_0 + c_1 q^{-1/2}$ for some (theory-dependent) constants $c_0, c_1$. 
	
After multiplying by $\sqrt q$, we get the following expression after applying the functional $\mathcal{F}_k$ to (\ref{eq:asdf}):
\begin{align}
		&\frac12\Lambda\left(\mbox{\large $\frac{c-1}2$}\right)\left(q^{\frac{1-c}2} - q^{\frac{c-1}2}\right)  - \sum_{\substack{i \in \mathcal{S} \\ i \neq \text{vac}}} \sum_{m=1}^\infty 2^{\frac{2-c}4} q^{\frac{c}4}  m^{\frac{c-2}2} E_i^{\frac{2-c}4} K_{\frac{c-2}2}\bigl(2 \pi m \sqrt{2q E_i}\bigr) \nonumber \\
		& +  \sum_{\substack{i \in \mathcal{S} \\ i \neq \text{vac}}} \sum_{m=1}^\infty 2^{\frac{2-c}4} q^{-\frac{c}4}  m^{\frac{c-2}2} E_i^{\frac{2-c}4} K_{\frac{c-2}2}\bigl(2\pi  m \sqrt{2E_i/q}\bigr) = -\frac{\pi \varepsilon_c}{12 \sqrt q} +  c_0 q^{1/2} + c_1,
  \label{eq:prefinalstuff}
\end{align}
where the function $K_{\frac{c-2}2}$ is a Bessel-K function. The LHS is manifestly anti-invariant under $q\leftrightarrow q^{-1}$ which forces $c_1 = 0$ and $c_0 = \frac{\pi \varepsilon_c}{12}$. Also we note that the first term in (\ref{eq:prefinalstuff}) equals the limit of the other two terms as $E_i\rightarrow 0$ (up to a divergent piece which becomes $q$-independent), so we can absorb it in $\mathcal{S}$. We therefore get
\begin{align}
		\sum_{i \in \mathcal{S}} \sum_{m=1}^\infty 2^{\frac{2-c}4} m^{\frac{c-2}2} E_i^{\frac{2-c}4} &\left(q^{-\frac{c}4}  K_{\frac{c-2}2}\bigl(2 \pi m \sqrt{2E_i/q}\bigr) - q^{\frac{c}4}  K_{\frac{c-2}2}\bigl(2 \pi m \sqrt{2q E_i}\bigr) \right) \nonumber \\ &~~~~~~~~~~~~~~~~~~~~~~~~~~~~~~~~~~~~~~~~~~~~= \frac{\pi \varepsilon_c}{12}\left(q^{1/2} - q^{-1/2} \right).
		\label{eq:finaleq}
\end{align}
	
The above derivation works in the Narain context, with $d=c$. We can actually generalize the derivation to generic CFTs, provided the similar crossing equation holds for scalar operators in the CFT spectrum. 
	
Let us take $c=0$ (i.e. $d = 0$) in (\ref{eq:finaleq}). The equation now looks like:
\begin{align}
		\frac{\sqrt 2}{q^{1/2} - q^{-1/2}} &\sum_{i \in \mathcal{S}} \sqrt{E_i} \sum_{m=1}^\infty \frac 1 m\left(K_1\left(2\pi m \sqrt{2 E_i/q}\right) - K_1\left(2\pi m\sqrt{2 E_i q}\right)\right) = \frac{\pi \varepsilon_0}{12}.
		\label{eq:epsilondef}
\end{align}
We find the integral of partition function over $\mathcal{F}$ only depends on the scalar spectrum of theory, which mimics the other term in the zero-wrapping sector that only includes the spinless particle excitations. Since we still need to integrate over the scalar spectrum, we need to discuss the convergent criteria of the above expression. The sum over $m$ converges rapidly, due to the exponential decay of the Bessel functions $K_1$ at large $m$. At large energy, however, the convergence is not as obvious due to the growth of the density of states of the scalar operators. At large $E_i$, the Bessel functions decay as
	\begin{align}
		\sum_{m=1}^\infty \frac 1 m\left(K_1\(2\pi m \sqrt{{2E_i}/q}\) - K_1\(2\pi m \sqrt{2 E_i q}\)\right) \sim e^{-2\sqrt2 \pi \sqrt{E_i} \sqrt{\text{min}(q, q^{-1}})}
	\end{align}
	On the other hand, the Cardy growth of scalars at large energy behaves as \cite{Cardy:1986ie} 
	\begin{align}
		\rho^{\text{scalars}}(E_i) \sim e^{2\pi \sqrt{\frac{c E_i}3}}
	\end{align}
Thus (\ref{eq:epsilondef}) converges if 
\bea
\label{eq:c6k}
	c \leq 6~ \text{min}(q, q^{-1}).
\eea
In the discussion, $q$ was introduced as an auxiliary parameter. So, physical quantities should be independent of $q$. Indeed, we numerically checked the identity (\ref{eq:epsilondef}) for various CFTs and various values of $q$'s that obey (\ref{eq:c6k}), and the resulting integral is always numerically $q$-independent to arbitrarily high precision. This allows us to take the limit when $q$ approaches 1 to remove the $q$-dependence. This gives:
	\begin{align}
		\varepsilon_0 = \frac{12\sqrt 2}{\pi} \sum_{i\in\mathcal{S}} \sqrt{E_i} \sum_{m=1}^\infty \left[\frac1m K_1(2\pi m \sqrt{2E_i} ) + 2\pi \sqrt{2E_i} K_0(2\pi m \sqrt{2E_i})\right].
		\label{eq:epsilonanalyticallycontinued}
	\end{align}
This expression converges for $c \leq 6.$
	
\section{More details on numerics}	
\label{sec:numerics}

In this Appendix we describe in more detail the numerical checks we have performed on the expression for $Z(\rho,\ssigma)$ given in equations (\ref{partfinal}), (\ref{finalone}) and (\ref{finaltwo}). 

The most practically troublesome terms in $Z(\rho,\ssigma)$ are the (finite number of) terms in $\mathcal{S}$ with $E_i < 0$ (meaning with scaling dimension less than $\frac c{12}$). If $E_i < 0$, the sums in (\ref{finalone}) and (\ref{finaltwo}) conditionally converge. Moreover the convergence is very slow. Indeed for $E_i < 0$, if we put some large cutoff $N$ in the sum over $m, n$ in (\ref{finalone}) and the sum over $m$ in (\ref{finaltwo}), and plot the truncated sums as a function of $N$, we see a wildly oscillating function whose envelope slowly decreases with $N$. In practice we find that the best way to estimate the final convergent sum is to compute the truncated sums for a large number of cutoffs, and then average the answers.

\def\twothree{\mbox{\large $\frac23$}}
For example, let us consider explicitly the $c=1$ self-dual free boson (i.e. the $SU(2)_1$ WZW model). The only state with $E < 0$ is the vacuum, which has $E = -\frac{1}{12}$. The two slow-converging sums we have to do are:
\begin{align}
    A(\rho) &\equiv \i \sqrt{\twothree}  \rho_2 \sum_{m=1}^{\infty} \left[\frac1m K_1\left(\i \pi \sqrt{\twothree} m\right) + \i \pi \sqrt{\twothree} K_0\left(\i \pi \sqrt{\twothree}m\right) \right] \nonumber \\
    B(\rho, \ssigma) &\equiv \i \sqrt{\frac{\rho_2 \ssigma_2}6} \sum_{(m,n)\neq(0,0)} \frac{1}{|m\ssigma+n|} K_1\left(\i \pi \sqrt{\frac{2\rho_2 |m\ssigma+n|^2}{3\ssigma_2}}\right).
    \label{eq:abdefs}
\end{align}
Both sums in (\ref{eq:abdefs}) are conditionally, but not absolutely, convergent. We find that practically speaking, averaging over different cutoffs gives more numerically accurate answers for (\ref{eq:abdefs}) than choosing one large cutoff. The remaining terms in (\ref{finaltwo}) converge rapidly and are easy to compute to very high precision.

As a check we have numerically verified the following nontrivial properties for a variety of CFTs with $c\leq 6$:
\begin{enumerate}
    \item $\hatF_{T\bar{T}}(\rho,\ssigma)$ is modular invariant in both $\rho$ and $\ssigma$
    \item If $\rho, \ssigma \in \i \mathbb R^+$, then $\hatF_{T\bar{T}}(\rho,\ssigma) \in \mathbb R + \frac{\i \pi n}2$ with $n \in \mathbb Z$ 
    \item $\hatF_{T\bar{T}}(\rho,\ssigma) = \hatF_{T\bar{T}}(\ssigma, \rho)$.
\end{enumerate}

\bibliographystyle{JHEP}
\bibliography{strongweak.bib}
\end{document}